\pdfoutput=1
\pdfmapfile{+mtpro2.map}
\documentclass[%
reprint, 
aps, prb,
superscriptaddress,
]{revtex4-2}

\usepackage{newtxtext}

\usepackage[subscriptcorrection]{mtpro2}
\pdfmapfile{+ ./texmf/fonts/map/dvips/mtpro2/mtpro2.map}

\usepackage{graphicx}
\usepackage{dcolumn}
\usepackage[super]{nth} 

\usepackage{hyperref}
\hypersetup{
    colorlinks,
    linkcolor={blue},
    citecolor={blue},
    urlcolor={blue}
}

\usepackage{physics}
\usepackage[version=3]{mhchem}

\usepackage{subfigure}

\usepackage[dvipsnames]{xcolor}

\usepackage{listings}
\usepackage{lipsum}
\usepackage{comment}

\usepackage[T1]{fontenc}

\newcommand*{\ii}{{\rm i}}
\newcommand*{\ee}{{\rm e}}

\newcommand*{\dg}{{\dagger}}

\newcommand{\veck}{{\mathbf{k}}}
\newcommand{\vecq}{{\mathbf{q}}}

\newcommand{\vecG}{{\mathbf{G}}}

\newcommand{\vecR}{{\mathbf{R}}}

\newcommand{\vecr}{{\mathbf{r}}}

\newcommand{\veczero}{{\mathbf{0}}}

\newcommand*{\abinitio}{{\textit{ab initio}} }

\newcommand{\uroman}[1]{\uppercase\expandafter{\romannumeral#1}}
\newcommand{\lroman}[1]{\romannumeral #1}

\begin{document}

\title{Theory of interaction-induced  charge order in CrSBr}

\author{Zhi-Hao Cui}
 \email{zhcui0408@gmail.com}
\affiliation{Department of Chemistry, Columbia University, New York, New York 10027, USA}

\author{Andrew J. Millis}
 \email{ajm2010@columbia.edu}
\affiliation{Department of Physics, Columbia University, New York, New York 10027, USA}
\affiliation{Center for Computational Quantum Physics, Flatiron Institute, 162 5th Avenue, New York, New York 10010, USA}

\author{David R. Reichman}
 \email{drr2103@columbia.edu}
\affiliation{Department of Chemistry, Columbia University, New York, New York 10027, USA}

\pdfbookmark[0]{Main Text}{Main Text}

\begin{abstract}
CrSBr is a layered van der Waals insulator with 
{a quasi one-dimensional electronic} structure  
{and} in-plane ferromagnetic
{order}.  Recent experimental work on Li-doped CrSBr reveals quasi-1D charge modulated states.
In this study, we develop \abinitio effective models for CrSBr 
to investigate these states and solve them using mean-field theory and density matrix embedding theory. The models are parametrized using density functional theory, the constrained random phase approximation, and the Rytova-Keldysh form of the long-range Coulomb interaction. Our simulations indicate the emergence of a charge density wave state characterized by cosine-like intra-chain density modulations and inter-chain phase shifts that minimize the Coulomb repulsion. Notably, at a doping level corresponding to $1/n$ electron per CrSBr unit, the most stable pattern exhibits a periodicity of $n$ cells, in agreement with experimental observations and Peierls’ instability arguments.
{Moreover, we demonstrate that the inter-chain order is sensitive to the range of Coulomb interactions. If the interaction {is hard-truncated to a short-ranged form}, some localized stripe-like states are computationally favored.}
This work provides an \abinitio framework for understanding the interplay of competing electronic and magnetic phases in CrSBr and related materials.
\end{abstract}

\maketitle

\section{Introduction}\label{sec:intro}

The unique electronic, optical, and mechanical properties of two-dimensional (2D) materials have sparked tremendous interest in the fields of condensed matter physics and materials science. Among these, 2D van der Waals (vdW) materials have attracted particular attention for their tunable magnetic and electronic properties, offering potential applications in spintronics, optoelectronics, and quantum devices~\cite{Ajayan16-2D-vdw-review,Duong17-2D-vdw-review,Burch18-2D-vdw-magnetism,Wang22-magnetic-2D-vdw-review}. One notable member of the 2D vdW family is CrSBr, a layered material with an anisotropic crystal structure and tunable magnetic order. Unlike 
electronically 2D materials such as graphene and \ce{MoS2}, CrSBr exhibits quasi-one-dimensional (quasi-1D) electronic behavior especially in the lowest-lying conduction bands, making it an intriguing platform for studying strongly correlated phenomena~\cite{Wilson21-CrSBr-interlayer,Cenker22-CrSBr-strain,Klein23-CrSBr-1D,Long23-CrSBr-ferromagnetic,Xie23-CrSBr-AFM-FM,Dirnberger23-CrSBr-magnon-polariton}.

A recent experimental study~\cite{Feuer24-CrSBr-doping} has demonstrated that CrSBr hosts a variety of correlated electronic phases when doped with electrons. Intercalation of the three-dimensional material with lithium changes the inter-layer magnetic coupling from antiferromagnetic to ferromagnetic while increasing the magnetic ordering temperature to approximately 200 K. Further, the electron doping {associated with}  the intercalation of CrSBr with lithium 
induces prominent charge-modulated phases.  {In the literature different terminologies, such as charge density wave (CDW) and Wigner crystal,   are used for charge modulated phases according to the amplitude and spatial structure of the charge modulations.} In this work, we refer to these modulated phases as CDW states although, {as will be seen}, the contrast between the minimum and maximum charge density may be large.
Given the rich physics observed in experiments,  a comprehensive understanding  of the interaction between electron doping, highly anisotropic band structure, and long-range Coulomb interactions 
in the doped CrSBr
is highly desirable.

Previous studies of 2D vdW materials have mostly relied on density functional theory (DFT) and its variants to predict the electronic structure. However, these methods alone are often insufficient for capturing correlation effects 
critical for describing phases like those observed in Ref.~\cite{Feuer24-CrSBr-doping}.  Moreover, in order to accommodate multiple CDW phases with varying wavelengths, a sufficiently large supercell and a dense $\veck$-point mesh are required, which poses significant challenges for DFT and other electronic structure approaches. Therefore, methods that combine an effective model Hamiltonian and many-body effects, such as quantum embedding theories~\cite{Georges96,Sun16QET,Gordon17-fragmentation-book,Ma21-quantum-impurity-embedding} are very useful for explicating the competition between different nearly degenerate electronic and magnetic states.

In this work, we develop \abinitio effective models for lithium-doped CrSBr and use them to explore the emergence of interaction-induced charge orders. We parametrize our model using Wannierized DFT bands in conjunction with the constrained random phase approximation (cRPA)~\cite{Aryasetiawan04,Miyake08,Vaugier12} to account for the Coulomb interactions. The quasi-1D nature of the system is captured by the Wannier orbitals-based Hamiltonian, and we study the competition between CDW, stripe, and uniform phases using both mean-field Hartree-Fock (HF) calculations and quantum embedding in the form of density matrix embedding theory (DMET)~\cite{Knizia12}). Our model not only provides insight into the ground-state behavior of CrSBr under electron doping but also sheds light on the role of long-range Coulomb interactions in stabilizing these phases.

The remainder of the paper is structured as follows: In Sec.~\ref{subsec:model}, we present the derivation of the effective models for electron-doped CrSBr based on DFT and cRPA calculations, and outline the computational methods used. Sec.~\ref{subsec:phases min} explores the phases within the minimal model and examines the impact of doping charges. In Sec.~\ref{subsec:parameter influence}, we examine the sensitivity of phase stability to the range of Coulomb interactions.  In Sec.~\ref{sec:conclusion}, we summarize our conclusions and propose directions for future research.

\section{Models and methods}\label{sec:model method}

\begin{figure*}[!htb]
\includegraphics[width=0.8\textwidth,  clip]{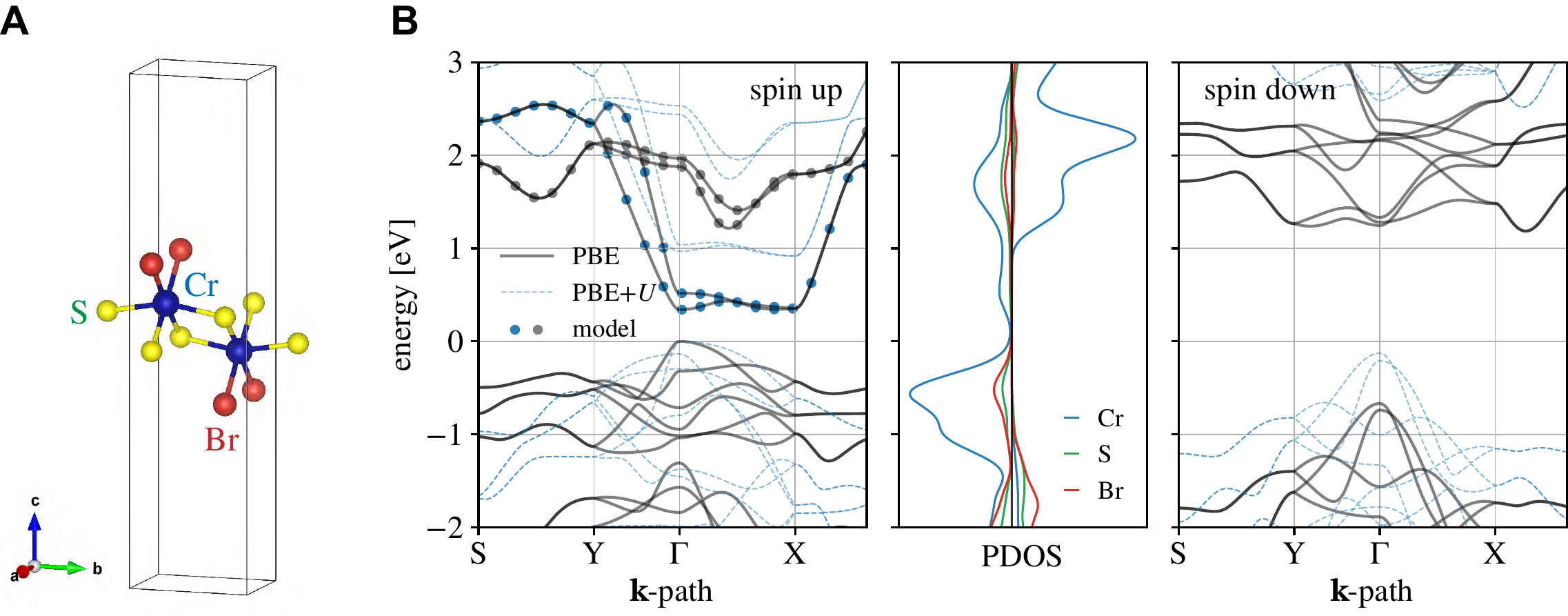}
\caption{{\bf Unit cell {(A) and}  band structure,and projected
density of states {(B)} of {ferromagnetic} undoped monolayer CrSBr.} The {bands for the} two spin channels are displayed in the left and right panels in (B) {and the spin-resolved density of states projected onto the different atomic orbitals is shown in the middle panel with spin up to the left and spin down to the right.} The bands calculated using the PBE functional are shown as solid lines (PBE+$U$ bands are shown as dashed lines), while those from the effective model are represented by dots (blue dots: $d_{z^2}$-like bands, grey dots: $d_{x^2-y^2}$-like bands). Special $\veck$ points in the Brillouin zone: $\Gamma$: (0, 0), X: ($\frac{1}{2}$, 0), Y: (0, $\frac{1}{2}$), S: ($\frac{1}{2}$, $\frac{1}{2}$). The energy zero is set to be the valence band maximum.}
\label{fig:band}
\end{figure*}

\subsection{Effective models from ab initio calculations}\label{subsec:model}

{Undoped, bulk} CrSBr consists of 2D planes held together by weak van der Waals forces {with ferromagnetic in-plane coupling} and antiferromagnetic coupling between the layers~\cite{Telford20-CrSBr,Lee21-CrSBr-magnetic,Rizzo22-CrSBr-magnetism,Bo23-CrSBr-magnetic,Xie23-CrSBr-AFM-FM,Shen24-exchange-interaction-CrSBr,Diederich24-exciton-CrSBr}.   Upon intercalation of lithium atoms accompanied by tetrahydrofuran (THF), the spacing between these layers increases 
{to a large value} ($\sim 13$ \AA), {so that} in lithium-doped CrSBr the interlayer interactions are negligible, allowing us to focus on a monolayer of CrSBr. {As shown in Fig.~\ref{fig:band} (A) the unit cell of a single 2D plane contains two  Cr atoms with different vertical ($c$) positions. Each Cr is approximately octahedrally coordinated (with local $C_{2v}$ symmetry) and basic formal valence considerations imply that in undoped CrSBr, each Cr is in the high-spin $d^3$ configuration with a half-filled fully polarized $t_{2g}$ shell and with the lowest unoccupied Cr orbitals the same-spin Cr $e_g$ d-orbitals.} {In panel (B) of} Fig.~\ref{fig:band}, we
{present} the band structure and {spin-resolved} density of states
{of monolayer CrSBr obtained} from DFT, using the Perdew-Burke-Ernzerhof (PBE) functional.  The system is found to be a ferromagnetic charge-transfer insulator {consistent with}
the
in-plane ferromagnetism {of the bulk material} 
~\cite{Telford20-CrSBr,Lee21-CrSBr-magnetic,Yang21-CrSBr-magnetic}). {The DFT band gap of $0.4$ eV is} substantially smaller that the experimental band gap $\sim 1.5$ eV. It is well known that the PBE functional can underestimate the band gap ~\cite{Bianchi23-CrSBr-gw-gap}) and that the underestimate can be partly remedied by adding a  local Hubbard $U$ and $J$ to the Cr $d$ orbitals. We also checked that PBE+$U$ calculation with $U = 3.68$ eV, $J = 0.39$ eV gives a band gap of 1 eV and a similar conduction band dispersion.   
{Consistent with the formal-valence considerations, the highest-lying} majority-spin valence bands around the Fermi level are hybrid bands from the Cr-$t_{2g}$ and Br, S ligand orbitals; whereas the {lowest-lying majority-spin} conduction bands are mainly composed of Cr-$e_{g}$-type orbitals with a minor contribution of S and Br. {The minority spin $d$-derived bands are much higher in energy}.

\begin{figure*}[!htb]
\includegraphics[width=0.9\textwidth,  clip]{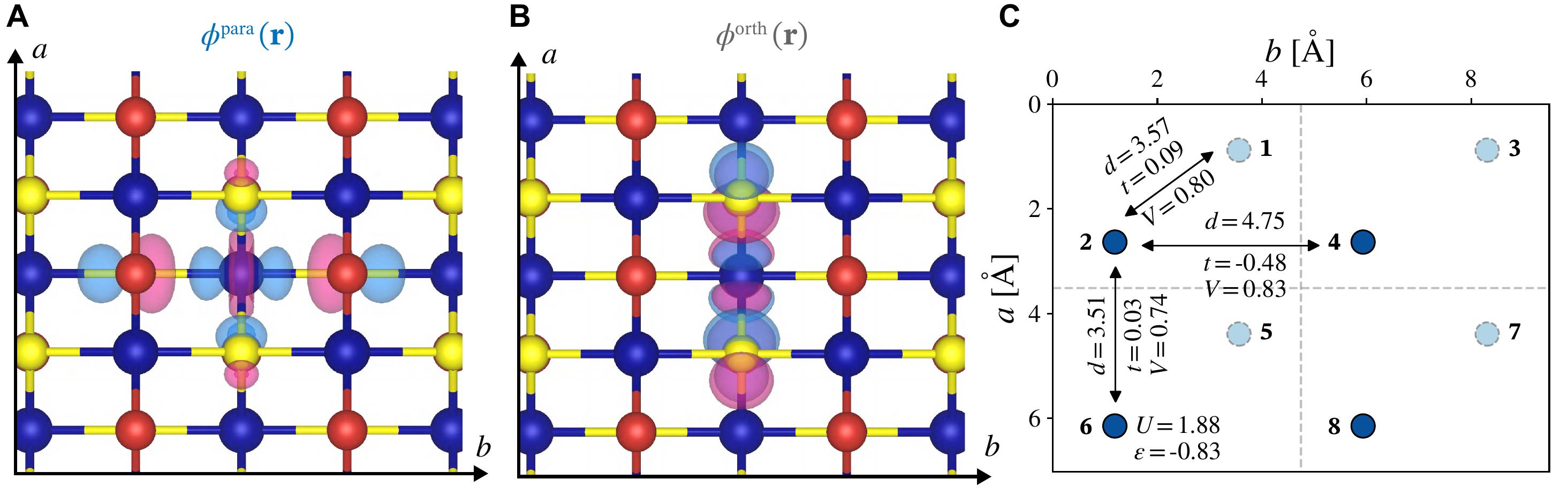}
\caption{{\bf $e_g$-like Wannier orbitals and representative parameters of model Hamiltonian.} 
(A) The Wannier orbital parallel to the $b$ direction, $\phi^{\text{para}}$, with a lower site energy $\varepsilon = -0.83$ eV (represents the $d_{z^2}$ bands well (blue dots in Fig.~\ref{fig:band}) and is used in the effective models). (B) The Wannier orbital orthogonal to the $b$ direction, $\phi^{\text{orth}}$, with a higher site energy $\varepsilon = -0.70$ eV (corresponds to the grey dots in Fig.~\ref{fig:band}). (C) Distance $d$, hopping $t$, and Coulomb interaction $V$ among $\phi^{\text{para}}$ (additional parameters are provided in Table~\ref{tab:parameter}). The chains with greater heights along the $c$ direction are shown in dark blue.
\label{fig:wann}}
\end{figure*}

{Of most interest for this work are the $e_g$-derived lowest-lying majority-spin conduction bands, which we expect to host electrons donated by the Li dopants. These are marked with}
 dots in Fig.~\ref{fig:band} (B). The two lowest conduction bands (blue dots) are almost flat along the $\Gamma$-X path {i.e. crystal $a$ direction}  but show significant dispersion along the $\Gamma$-Y direction {(crystal $b$ direction), indicating a quasi-one-dimensional chain structure.}
{The two lowest} bands can be localized as Wannier orbitals aligned parallel to the $b$ direction [Fig.~\ref{fig:wann} (A)]. The other two bands (grey dots), situated between 1.0 and 2.2 eV, consist of Wannier orbitals oriented perpendicular to the chain direction [Fig.~\ref{fig:wann} (B)]. These orbitals all display an $e_g$-like character and are centered on the Cr atoms. However, the orbital isosurface plot reveals non-negligible orbital spreading around the S and Br ligands, leading to non-trivial long-range hybridization   
in the subsequent simulations.

We use the spin-up Wannier orbitals to construct effective models for the monolayer CrSBr. {The one-body parameters are obtained by projecting the DFT Kohn-Sham Hamiltonian onto the Wannier orbitals. The Coulomb interactions are calculated using the constrained random phase approximation (cRPA).}, as shown in  Fig.~\ref{fig:wann} (C). 
The overall Hamiltonian can be written as,
\begin{equation}
\begin{split}
\mathcal{H} = \sum_{\vecR ij} t^{\vecR}_{ij} a^{\dg}_{\vecR i} a_{\veczero j} + \frac{1}{2} \sum_{\vecR ij}  V^{\vecR}_{iijj} n_{\vecR i} n_{\veczero j} ,
\end{split}    
\end{equation}
where $a^{(\dg)}_{\vecR i}$ is the annihilation (creation) operator for spinless electrons and $n_i$ is the corresponding density operator $n_{\vecR i} = a^{\dg}_{\vecR i} a_{\vecR i}$ of Wannier orbital $\phi_{i}$ at cell $\vecR$. For the \textit{full model} used in this work, the orbital indices are restricted to the lowest two bands (this is valid as the doping concentration is not large) and the one-body terms $t$ include all site energies and hopping parameters. The two-body Coulomb interactions $V$ contain only the density-density interactions {because the exchange interactions, most notably the on-site Hunds coupling, are saturated in the assumed ferromagnetic state. Moreover, because we consider only very low doping of the conduction band, double counting corrections are not important}. 

\begin{table}[b]
\caption{\label{tab:parameter}
Model hopping $t$ and Coulomb $V$ parameters. The distance $d$ and orbital ordering can be found in Fig.~\ref{fig:wann} (C). The underlined values are used in the minimal model. The hopping parameters beyond the $2 \times 2$ cell are not shown, but included in the full model.}
\begin{ruledtabular}
\begin{tabular}{cccccccc}
$i$ & $j$ &$d_{ij}$ [\AA] & $t_{ij}$ [eV] & $V_{iijj}$ [eV] & comment\\
\hline
2 & 2  & 0.00 & -0.831 & 1.875 & on-site \\
2 & 6  & 3.513 & 0.034  & {\underline{\bf{0.742}}} & \nth{2}-neighbor inter-chain \\
2 & 1  & 3.571 & 0.093 & {\underline{\bf{0.804}}}  & \nth{1}-neighbor inter-chain \\
2 & 4  & 4.746 & {\underline{\bf{-0.481}}} & {\underline{\bf{0.834}}} & intra-chain \\
2 & 8  & 5.904 & 0.006  & 0.634 &  \\
1 & 6  & 6.118 & -0.068 & 0.608 & \\
2 & 3  & 7.602 & 0.002        & 0.631 & \\
3  & 6     & 9.081 & 0.000             & 0.542 & \\
$\cdots$ \\
\end{tabular}
\end{ruledtabular}
\end{table}
Representative parameters are shown in Fig.~\ref{fig:wann} (C). {We see that the electron hybridization has an interesting structure. The upper (lower) Cr atom in one unit cell is relatively strongly hybridized with the upper (lower) Cr atoms in the unit cell adjacent along the $b$ direction (intra-chain hopping $0.48$ eV) and is much weaker coupled to each other or in the $a$ direction (inter-chain hopping 0.09 and 0.03 eV). This implies that the low-lying electronic structure of the conduction band may be understood as two weakly coupled chains running along the crystal $b$ direction, one corresponding to the upper and one to the lower Cr atom in the unit cell. }
The on-site Coulomb interactions are approximately 2 eV, which is lower than the commonly used value for Cr $d$ orbitals ($\sim 4$ eV), due to the inclusion of on-site screening effects from the $t_{2g}$ orbitals. {Although the hopping parameters are dominant along the chain direction, the Coulomb repulsions remain relatively long-range in both directions. } 
Due to the extended tails of the Wannier orbitals, the intra-chain Coulomb repulsion ($0.83$ eV) is larger than the inter-chain values ($0.80$ and $0.74$ eV), even though the intra-chain distance is greater ($4.75$ \AA). The complete set of parameters can be found in Table~\ref{tab:parameter}. {In addition to the interactions shown, we also include in our model} long-range interactions beyond the $2 \times 2$ cell by using a Rytova-Keldysh model (see Sec.~\ref{subsec:parameter influence} for details).

{In addition to the full model above, we consider a \textit{minimal} model }
\begin{equation}\label{eq: min model}
\begin{split}
    \mathcal{H}^{\text{min}} =& t^{\text{intra}} \sum_{i} a^{\dg}_{i} a_{i+r_b} + \frac{1}{2} V^{\text{intra}} \sum_{i} n_{i}n_{i+r_b} \\
    &+ \frac{1}{2} V^{\text{inter}} \sum_{i} n_{i}n_{i+r_{ab}} + \frac{1}{2} V^{\text{inter}'} \sum_{i} n_{i}n_{i+r_a}
\end{split}
\end{equation}
{that only contains four parameters: the intra-chain hopping $t$ [e.g., between orbitals 2 and 4 in Fig.~\ref{fig:wann} (C)] and the intra- and inter-chain Coulomb interaction $V$ [the pair interactions in Fig.~\ref{fig:wann} (C)]. The vector $r_a$ ($r_b$) connects the neighboring sites in the $a$ ($b$) direction, while $r_{ab}$ connects the two Cr atoms in the unit cell. } 

\subsection{Mean-field and density matrix embedding theory }\label{subsec:mf dmet}

With the two models
, we can carry out mean-field Hartree-Fock (HF) calculations as well as density-matrix embedding theory (DMET) for the doped systems. DMET provides a natural framework to combine mean-field approaches with electron correlations. The technical details and implementation of DMET can be found in  Refs.~\cite{Wouters16,Cui20-dmet-solid,Cui21-cuprate-parent-state}. In this work, we employ a coupled-cluster (CC)~\cite{Shavitt09book} in Hartree-Fock (HF) embedding, meaning that CC is used to handle electron correlations within the supercell, while HF is used to treat the longer-range interactions

In experiments, Li-THF intercalates between the layers of CrSBr and effectively introduces electrons into the layers. We therefore directly add electrons into the models (i.e., rigid-band approximation). To simulate different orders, we use a rectangular supercell of $1 \times l_{\text{sc}}$, where $l_{\text{sc}}$ is the cell length along the $b$ direction (or $2 \times l_{\text{sc}}$ cell if the modulation along $a$ direction is considered). The whole lattice is tiled by such supercells and is large enough to accommodate different doping concentrations.


The DMET calculations are based on the converged HF solutions, with the supercell chosen as the impurity. The embedding problem, which comprises the impurity and its surrounding bath, is addressed using a coupled-cluster singles and doubles (CCSD) solver.

\subsection{Computational details}\label{subsec:details}

The crystal structure of CrSBr was taken from Ref.~\cite{Lopez22-CrSBr-struct} ($a = 3.5127$ \AA \, and $b = 4.7458$ \AA). A 20 \AA \, vacuum was added to the monolayer in the $c$-axis to avoid artificial interactions between periodic images.

The DFT calculations using both the PBE functional~\cite{Perdew96PBE} and PBE$+U$ (in the rotationally invariant formalism~\cite{Lichtenstein95}, with $U = 3.68$ and $J = 0.39$ eV from Ref.~\cite{Bianchi23-CrSBr-lifshitz-transition}; other $U$, $J$ values can be found in Ref.~\cite{Rudenko23-CrSBr-dielectric}) were carried out with the \textsc{VASP} package~\cite{Kresse94-vasp, Kresse96-vasp, Kresse96}. A plane wave cutoff energy of 500 eV was employed, and the Brillouin zone of the 2-atom unit cell was sampled with a 
$\veck$ mesh of $18 \times 14 \times 1$. {Maximally localized Wannier functions~\cite{Marzari12RMP} for the two spin channels were generated for all four lowest conduction bands (rather than the lowest two, to ensure smooth disentanglement), using the \textsc{wannier90} code~\cite{Mostofi14}.}
The cRPA calculations using the projector method~\cite{Kaltak15-cRPA-thesis} in \textsc{VASP}, employed the PBE band structures and Wannier orbitals in a $2 \times 2$ supercell with a $9 \times 7 \times 1$ $\veck$ mesh (using 16 target Wannier orbitals with a total number of bands of 512). The long-range (real-space) screened Coulomb potential was calculated via the $G_0W_0$ procedure~\cite{Hedin65,Hybertsen86,Aryasetiawan98RPP,Shishkin06}.

For simulating different phases, the supercell size 
$l_{\text{sc}}$ ranged from 4 to 15, with an additional 
$\veck$ mesh of $42 \times 20 \times 1$ to ensure adequate sampling of the supercell Brillouin zone. The HF and DMET calculations were performed using the \textsc{libDMET} package~\cite{Cui20-dmet-solid,Cui21-cuprate-parent-state,Cui23-cuprate-doping}. The DMET was computed in a 1-shot manner (i.e., without self-consistency) due to the relatively weak Coulomb interactions. The interacting bath formalism~\cite{Wouters16} was applied, and the long-range Coulomb interaction was treated using the mean-field theory. The spin-unrestricted CCSD solver was adapted from the \textsc{PySCF} package~\cite{Sun18pyscf, Sun20pyscf}.

\section{Results}\label{sec:results}

\subsection{Phase diagram of CrSBr minimal model}\label{subsec:phases min}

\begin{figure}[!htb]
\includegraphics[width=0.49\textwidth,  clip]{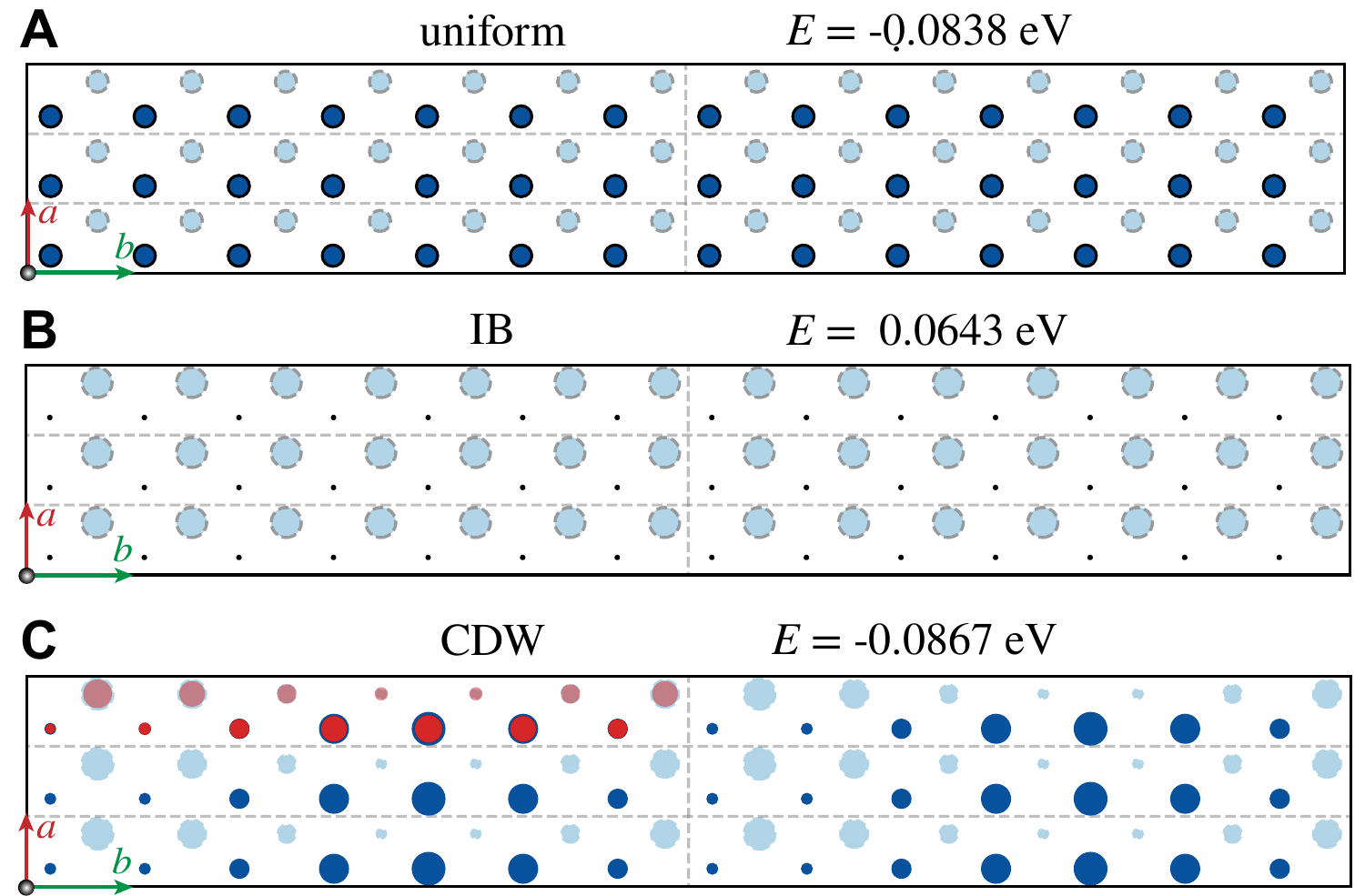}
\caption{{\bf Phases of electron-doped CrSBr using the minimal model (with doping concentration $1/7$ $e$ per CrSBr).} (A) Uniform phase, (B) inversion-symmetry broken (IB) phase, and (C) charge density wave (CDW) phase. In these diagrams, the dark and light blue dots represent two distinct types of Cr sites, each at different heights along the 
$c$-axis. The size of the circles corresponds to the doping density at each site. The dashed line indicates the supercell used in the calculations. The CDW state density from mean-field HF (blue circles) and that from correlated DMET calculation (red circles) are compared in the first supercell. 
\label{fig:phases}}
\end{figure}

\begin{figure}[!htb]
\includegraphics[width=0.49\textwidth,  clip]{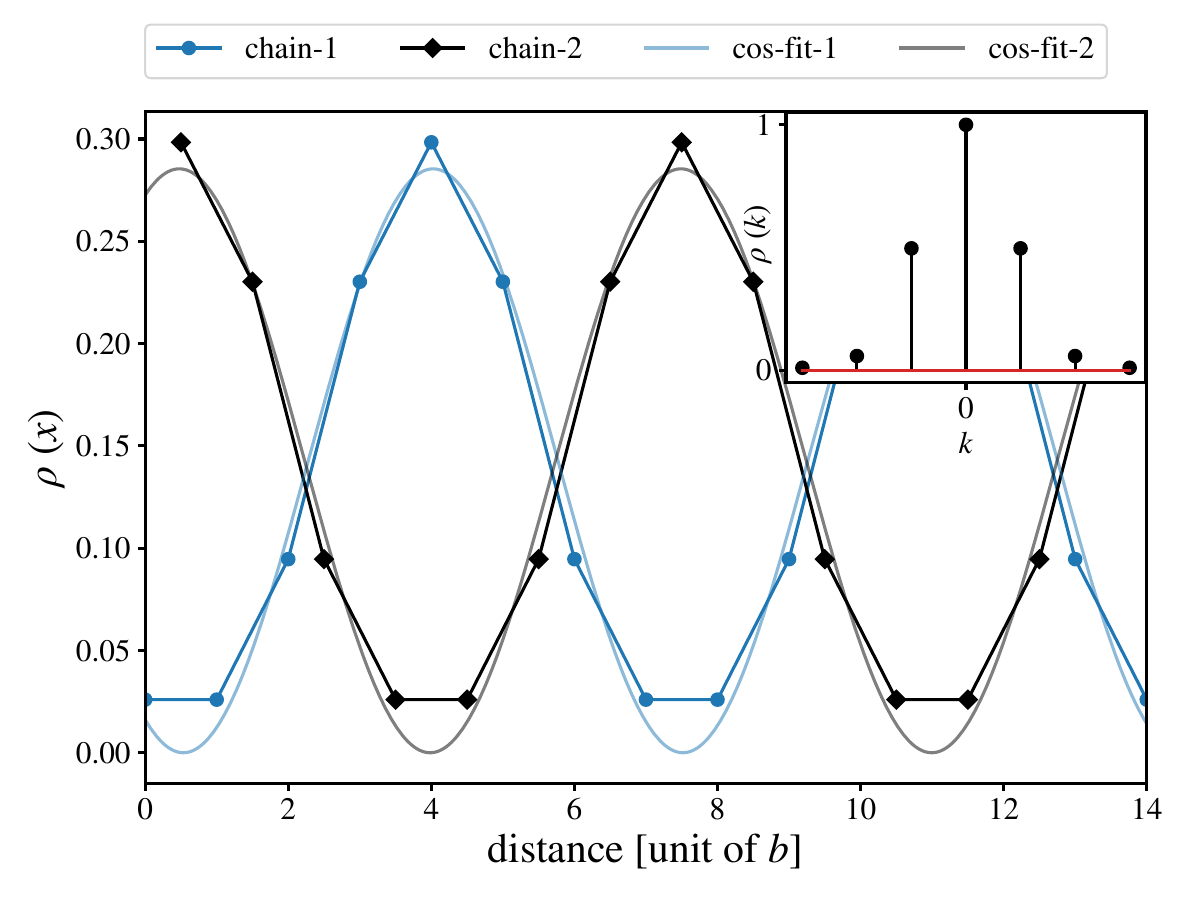}
\caption{Electron density of the CDW state (from the minimal model) varies with distance along $b$ direction. The densities in the two chains are fitted to cosine waves Eq.~\eqref{eq: cos fit} (with parameters $C = 0.143$, $\lambda = 6.98$, $N = 2$). The inset shows the Fourier components of $\rho$.
\label{fig:cos wave}}
\end{figure}

We start with a discussion of the ground state of the effective minimal model 
of CrSBr. {The model has three relevant symmetries: discrete translation along the $a$ and $b$ directions and an inversion symmetry that interchanges the Wannier orbitals corresponding to the upper and lower Cr atoms.} Several potential phases can be stabilized in this system (Fig.~\ref{fig:phases}): (a) uniform, where charges are evenly distributed and none of the symmetries of the model are broken; (b) inversion-symmetry broken (IB) state 
, where occupied and unoccupied chains alternate {but within each chain the charge is uniformly distributed (so that translation symmetry is preserved); and (c) charge density wave (CDW), characterized by {broken translational symmetry along the $b$-axis and broken inversion symmetry, but with the translation symmetry along $a$  unbroken}. 

{Fig.~\ref{fig:cos wave} presents the structure of} the CDW phase {in more detail}, showing that  the electron density {has a period-7 variation in space along $b$ when doping $x = 1/7$ $e$ per CrSBr.  One notable feature is the pronounced modulation in electron density between sites. In the Hartree-Fock approximation, the largest local density reaches about $0.3e$, while the smallest density is nearly zero. When electron correlation is introduced via quantum embedding theory [Fig.~\ref{fig:phases} (C)], this density difference decreases only slightly (with the largest change 0.04 $e$). In general, electron correlation tends to reduce the localization of charges, though the effect is marginal here due to the relatively small interactions ($U/t \sim 3.9$ and $V/t \sim 1.7$) and the large CDW gap (in mean-field theory, the gap is approximately 0.26 eV). Consequently, the CDW phase exhibits significant characteristics of a Wigner crystal~\cite{Wigner34-wigner-crystal,Ung23-wigner-crystal-moire,Tsui24-wigner-crystal-expt}, with well-separated localized charge regions driven purely by electron-electron Coulomb repulsion. We also see that although the amplitude of the density modulation is very large, the density profile is nearly cosinusoidal, with the \nth{0} and \nth{1} Fourier components much larger than the higher harmonics. } For a purely 1D spin-polarized system, the Fermi vector is given by
\begin{equation}
    k_{\text{F}} = \frac{\pi n_{\text{nelec}}}{l_b} ,
\end{equation}
where $n_{\text{nelec}}$ is the number of electrons per site and $l_b$ is the spacing between the sites in the chain direction {and the Peierls mechanism of ordering via a Fermi surface instability predicts that the ordering wave vector is $Q = 2k_{\text{F}}$ ~\cite{Gruner88-cdw-review,Giuliani08-quantum-liquid-book,Overhauser62-SDW}. Even though the Peierls argument applies for weak coupling (small amplitude modulation), the amplitude modulation is very large here. The periodicity $\lambda$ is predicted to be ${\pi}/{k_{\text{F}}} = l_b / n_{\text{nelec}}$.
In Fig.~\ref{fig:cos wave}, $n_{\text{nelec}} = 1/7 e$ and therefore the periodicity spans 7 unit cells. }

We next consider the variation transverse to $b$. It is useful to adopt a notation that ignores the two-atom unit cell structure and indexes the chains sequentially by an index $n$. Ignoring the higher harmonics we may then write for the charge density as a function of chain index $n$ and coordinate $x$ along $b$, as  
\begin{equation}\label{eq: cos fit}
    \rho_n (x) = C\qty[\cos(\frac{2\pi}{\lambda} x + \frac{2n\pi}{N}) + 1] 
\end{equation}
where $\lambda$ corresponds to the wave periodicity along the $b$ direction (parallel to the chain) and $N$ represents the periodicity in $a$ direction, meaning that the phase shift of chain $n$ is identical to that of chain $(n+N)$.   Here $N=2$ describes a situation in which the two neighboring chains exhibit a phase shift of $\pi$. This phase shift minimizes the inter-chain Coulomb repulsion (see below for further discussion), and the energy gain renders the CDW state energetically more favorable than the uniform state. This pattern aligns well with experimental STM observations~\cite{Feuer24-CrSBr-doping}. 
{In the STM measurement, only the top layer charge modulation is detected and its periodic pattern matches the dark blue atoms in Fig.~\ref{fig:phases}(C), when doping is approximately $1/7 e$ per CrSBr.}


The inversion-symmetry broken (IB) state represents another stripe-like charge order,
where one chain is fully occupied and the adjacent chain is completely empty. This configuration entirely eliminates the nearest-neighbor chain Coulomb repulsion. However, it simultaneously increases the intra-chain repulsion and enhances the second-neighbor inter-chain repulsion, which will destabilize the phase  (see Sec.~\ref{subsec:parameter influence} for more discussion).



\begin{figure}[!htb]
\includegraphics[width=0.49\textwidth,  clip]{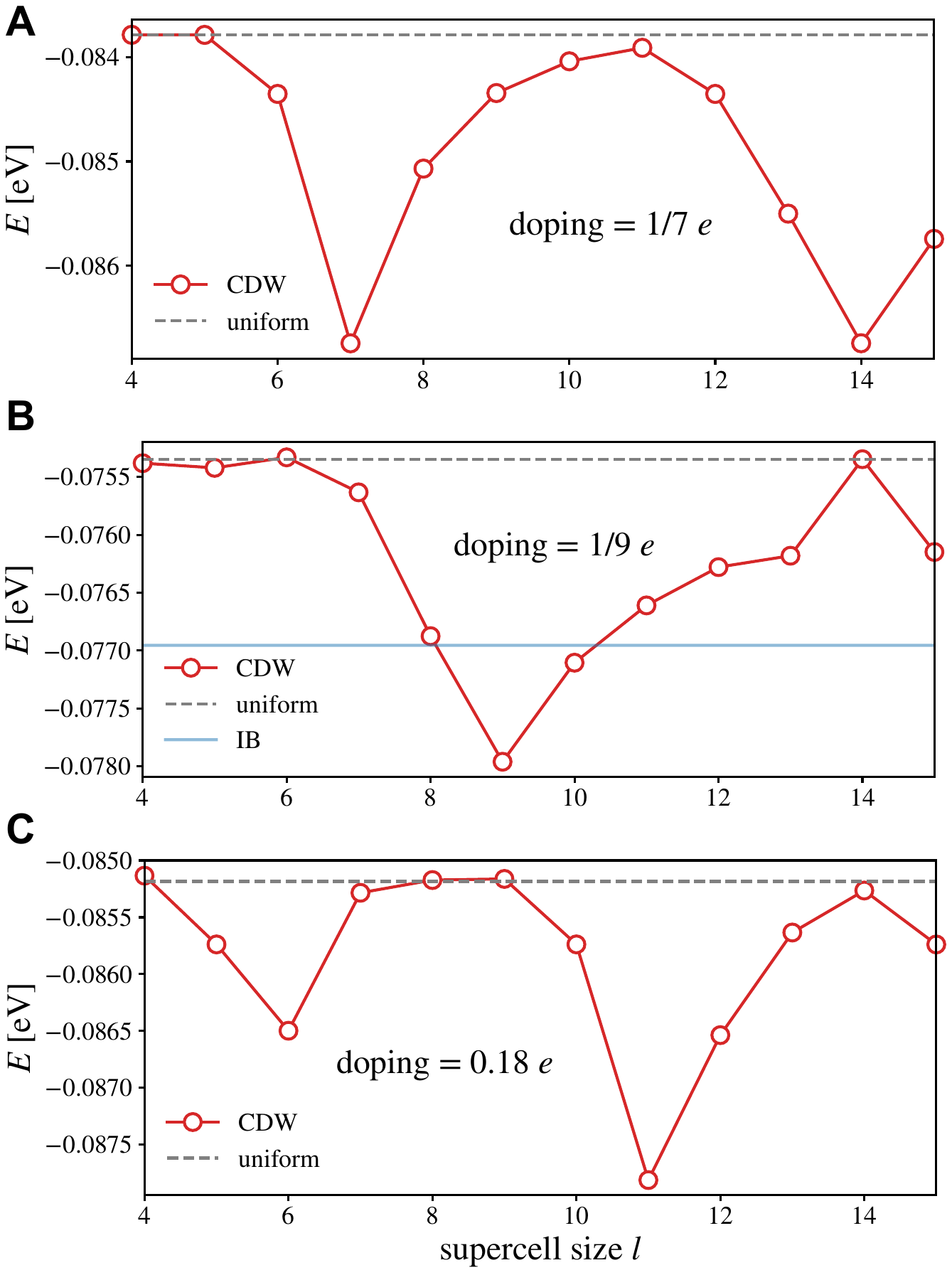}
\caption{{\bf Phase diagram at different doping concentrations as a function of supercell size (periodicity).} (A) $x = 1/7 e$, (B) $x = 1/9 e$, (C) incommensurate doping $x = 0.18 e$ per Cr.
\label{fig:doping vs l}}
\end{figure}

Since we employ real-space-based simulations, it is worth checking the dependence of ground-state energy $E$ on the computational supercell size $l$.
In Fig.~\ref{fig:doping vs l}(A) and (B), we show the energy curves for $1/7 e$ and $1/9 e$ doping concentrations. These doping levels exhibit energy minima at exactly 7 and 9 multiples of the unit cell length, respectively, which matches the prediction of $\lambda = l_b / n_{\text{nelec}}$. The experimental CDW wavelength can then be expected to vary based on different doping concentrations (e.g., controlled through intercalation, or through electrolyte gating that allows for continuous doping changes). 
When doping concentration does not fit into the computational supercell (i.e., the number of electrons per supercell is no longer an integer), the energy curve can exhibit multiple local minima. As shown in Fig.~\ref{fig:doping vs l} (C), the $0.18e$ doping leads to minima at 6, 11, and other values, which are close to multiples of 5.55. Thus for our analysis, a sufficiently large supercell is needed to accommodate the ideal periodicity and avoid boundary artifacts.

\subsection{Sensitivity to truncation of  interactions}\label{subsec:parameter influence}




In the previous section, we used a minimal model to understand the basic physics of density wave ordering. Here we show that the structure of the ordered state is in fact surprisingly sensitive to the range of the inter-particle interaction. We begin by showing that the addition of a next-neighbor repulsion of appropriate strength can change the periodicity transverse to the stripes. We then show that various truncations of the physical long ranged Coulomb interaction can destroy the CDW order altogether, but that use of the physical untruncated Coulomb interaction restores the CDW order discussed in the previous section.

We first consider the influence of the next-nearest neighbor Coulomb repulsion ($V'$). From Eq.~\eqref{eq: cos fit}, the local energy density at a continuous position $x$ can be expressed as
\begin{equation}
    e[N, x] = \frac{1}{N} \sum^{N-1}_{n=0} \qty[4 V \rho^{N}_{n} (x) \rho^{N}_{n+1} (x) + 2 V' \rho^N_n (x) \rho^N_{n+2} (x)],
\end{equation}
where $4$ and $2$ are the coordination numbers.
After integration over $x$, we have the total energy as a function of $N$ (the periodicity in the $a$ direction), \begin{equation}\label{eq:energy classical}
    E [N] = \lambda C^2  \qty[4 V + 2 V' + 2 V \cos(\frac{2\pi}{N}) + V' \cos(\frac{4 \pi}{N})] .
\end{equation}
It is clear if the nearest neighbor repulsion dominates, to minimize the energy, the third term would require $N = 2$ and the phase shift is $\pi$. On the other hand, if $V'$ dominates, the last term would require $N = 4$, which gives a $\pi / 2$ shift, sliding from one chain to another [Fig.~\ref{fig:vprime} (A)].
\begin{figure}[!htb]
\includegraphics[width=0.49\textwidth,  clip]{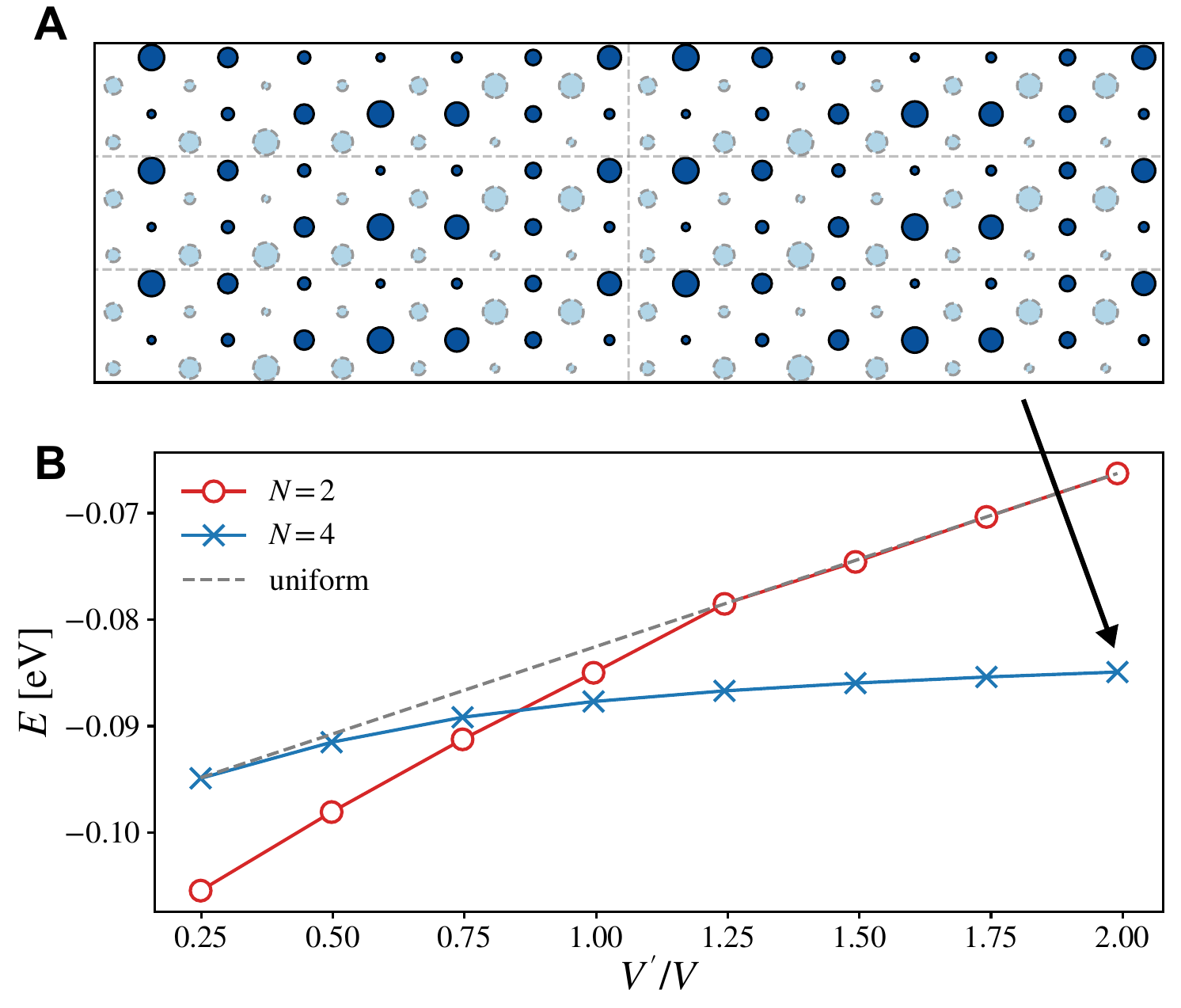}
\caption{(A) CDW state with a phase shift of $\pi / 2$ ($N = 4$) at $V' / V = 2.0$. (B) Energies of $N = 2$ ($\pi$ shift) and $N = 4$ ($\pi / 2$ shift) CDW phases as a function of ratio $V' / V$ (next-nearest / nearest inter-chain coupling).
\label{fig:vprime}}
\end{figure}
The energy curves are shown in Fig.~\ref{fig:vprime} (B), where the transition from $N = 2$ to $N = 4$ is predicted to occur at $V' / V = 1$, as derived from Eq.~\eqref{eq:energy classical}. In the actual calculations, the transition point is approximately $0.86$, which is in reasonable agreement with the above classical prediction $1.0$. Additionally, the uniform phase consistently has a higher energy than the various CDW states across the parameter range. These findings indicate the sensitivity of phases to Coulomb interactions, and therefore we consider the full interactions as follows.




\begin{figure}[!htb]
\includegraphics[width=0.49\textwidth,  clip]{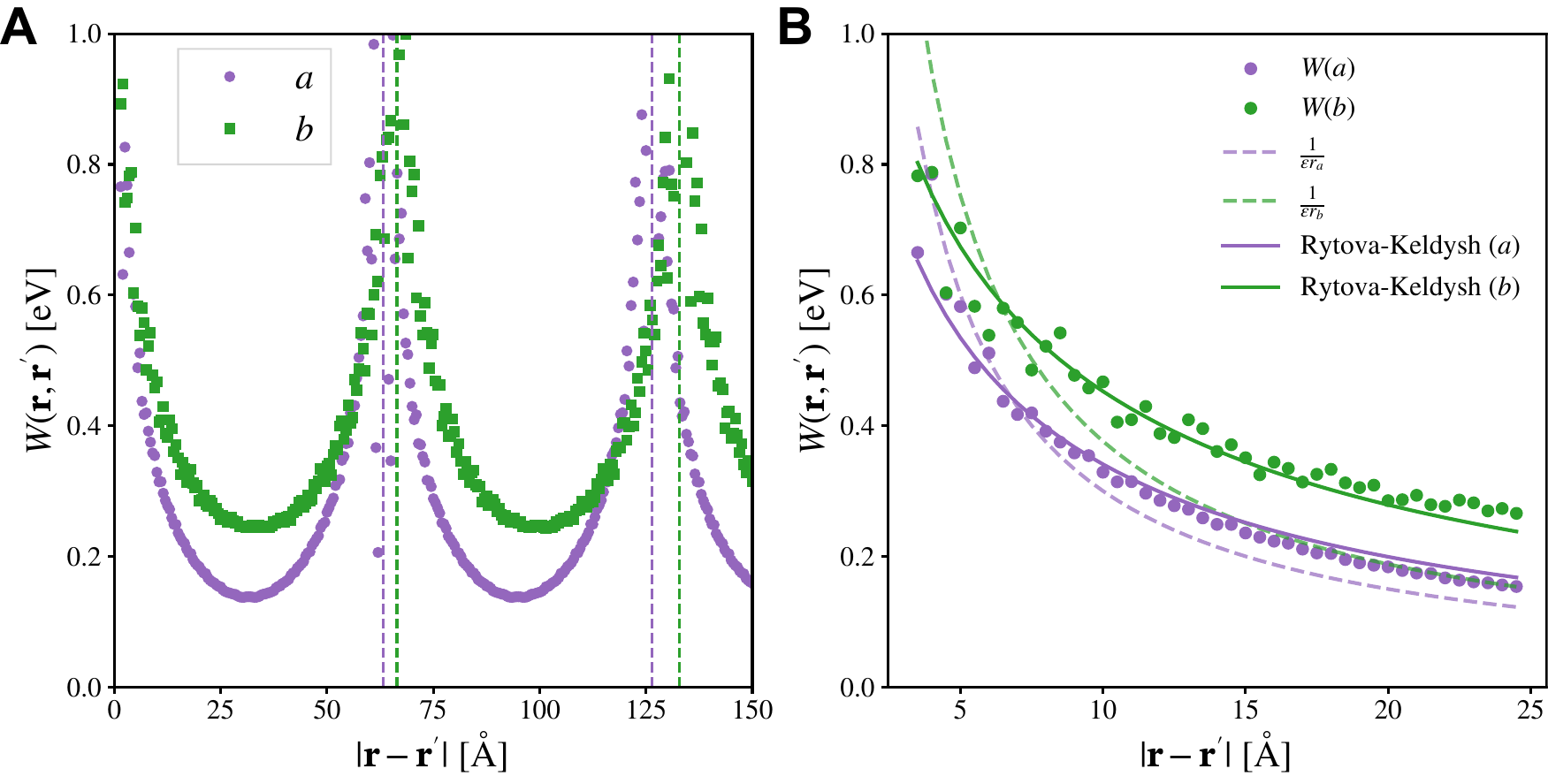}
\caption{(A) Screened Coulomb potential $W$ in real space $\vecr$ (along the $a$ and $b$ directions). The dashed lines represent the Born–von Karman boundary conditions. (B) $W$ is fitted using dielectric models: the $1 / \varepsilon r$ form and the Rytova-Keldysh form, with parameters $A = 0.7045$ eV, $B_x = 0.0989 \text{\AA}^{-1}$, and $B_y = 0.0650 \text{\AA}^{-1}$.
\label{fig:long range}}
\end{figure}
As shown in Table~\ref{tab:parameter}, the screened Coulomb interactions between Wannier functions decay slowly. One approach to account for the long-range effects is to estimate the screened Coulomb interaction in real space:
\begin{equation}
    W_{\vecG \vecG'}(\vecq) = 4\pi  \frac{1}{|\vecq+\vecG|} \varepsilon^{-1}_{\vecG \vecG'}(\vecq, \omega=0)  \frac{1}{|\vecq+\vecG'|},
\end{equation}
\begin{equation}
    W(\vecr, \vecr') = \sum_{\vecq \vecG \vecG'} \ee^{-\ii(\vecq+\vecG)\cdot \vecr} W_{\vecG\vecG'}(\vecq) \ee^{\ii(\vecq+\vecG')\vecr'} ,
\end{equation}
where $\vecG, \vecG'$ are plane-wave basis and $\vecq$ are the vectors sampling the first Brillouin zone, frequency $\omega$ is taken as 0 (static limit) for simplicity.
From Fig.~\ref{fig:long range} (A), the screened Coulomb interaction is anisotropic and decays more slowly in the $b$ direction (along the chain). The Coulomb potential of a 2D film (in the vacuum or the substrate's dielectric constant is small) can then be fitted to constant screened potential,
\begin{equation}
    W(r_x, r_y) = \frac{1}{\sqrt{\varepsilon^2_x r^2_x + \varepsilon^2_y r^2_y}} ,
\end{equation}
or the Rytova-Keldysh form~\cite{Rytova67-screened-coulomb,Keldysh79-screened-coulomb,Berkelbach13-dichalcogenides},
\begin{equation}\label{eq: RK potential}
    W(r_x, r_y) = A \qty[H_0 \qty(\bar{r}) - Y_0 \qty(\bar{r})] ,
\end{equation}
where $H_0$ and $Y_0$ are the Struve function and the Bessel function of the second kind, respectively. The asymptotic behavior of such potential is $\propto -\ln(r)$ in a short distance, and is $\propto 1/\varepsilon r$ in a long distance. We have generalized the original formulation to anisotropic cases by introducing the effective distance $\bar{r} \equiv \sqrt{B^2_x r^2_x + B^2_y r^2_y}$, with $A$, $B_x$, and $B_y$ as fitting parameters. 
The fitting curves and fitted parameters are shown in Fig.~\ref{fig:long range} (B). It is evident that for a 2D system like monolayer CrSBr, the simple constant screening model fails to reproduce the long-range behavior and decays much faster, whereas the Rytova-Keldysh model with anisotropic parameters matches the \abinitio data points well. We then use this model to compute the long-range Coulomb matrix elements beyond the $2 \times 2$ supercell. One can in principle fit a different Rytova-Keldysh potential for a half-infinite surface system (to mimic the STM experiments or different dielectric substrates). However, due to the large spacing between the CrSBr layers, we used the single-layer fit. 

\begin{figure*}[!htb]
\includegraphics[width=0.99\textwidth,  clip]{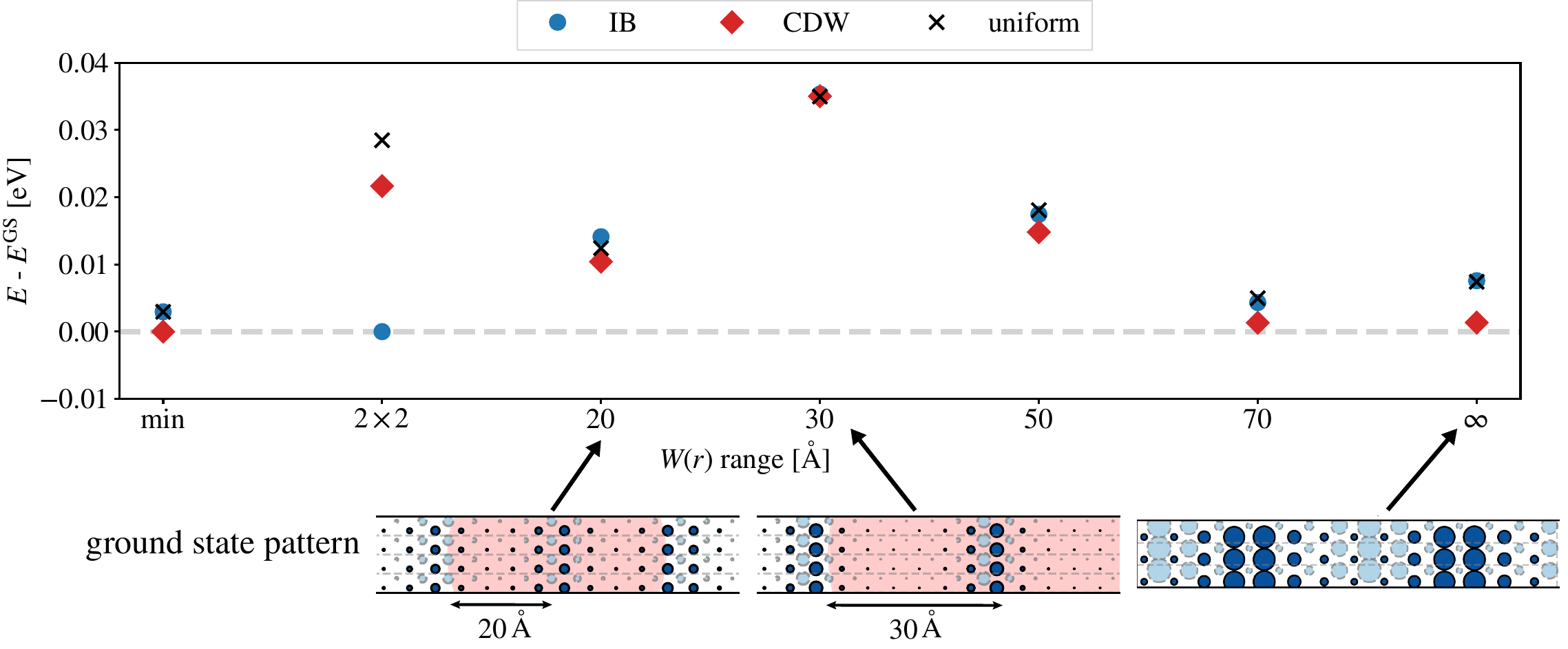}
\caption{Phase diagram as a function of Coulomb length truncation. The ground-state energy $E_{\text{GS}}$ is set to zero for different Coulomb ranges. Some representative ground-state patterns (different from the CDW, uniform, and IB phases discussed before) are also shown.
$2\times 2$ means the Coulomb elements are limited to the values in Table~\ref{tab:parameter}. The red area indicates the boundary (radius) of Coulomb truncations (for a central Cr site), and $\infty$ means the long-range interactions are not truncated.
\label{fig:phase vs coulomb}}
\end{figure*}
We gradually include more Coulomb interactions, ranging from short-range to long-range (see Fig.~\ref{fig:phase vs coulomb}). Initially, we incorporate all $t_{ij}$ terms across the entire lattice, and all $V_{iijj}$ interactions within the $2\times2$ supercell (as shown in Table~\ref{tab:parameter}). Interestingly, this results in the inversion-symmetry broken phase as the ground state. This occurs because inter-chain interactions beyond the $2\times 2$ cell are not included. For instance, the interactions between the \nth{1} and \nth{5} chains (which are separated by a distance of only $d = 7.025$ \AA) are neglected. When such Coulomb parameters are incorporated, the IB phase is heavily penalized and becomes unstable across all ranges in Fig.~\ref{fig:phase vs coulomb} beyond the $2 \times 2$ cell. In fact, except for the $2 \times 2$ and $20$ \AA \, truncation cases, all phases initially labeled as IB spontaneously converge to the uniform phase.

When the Rytova-Keldysh type interaction $W(r)$ is hard-truncated at medium ranges (20–50 \AA), the ground state evolves into localized stripes along the $a$ direction. The distance between the edges of these stripes is approximately equal to the truncation length, which minimizes the Coulomb interactions from the medium-range tails of the potential. As the truncation length increases significantly (e.g., 70 \AA), the CDW state re-emerges as the ground state, with a slight modification in charge separation (and small symmetry breaking between the two chains, originated from the slight different $\veck_{\text{F}}$'s for the two quasi-1D systems), compared to the CDW state in the minimal model. 

Finally, we examine the case without any truncation, where all possible Coulomb pair interactions ($V_{iijj}$) are included, as described by model Eq.~\eqref{eq: RK potential}. To keep neutrality in simulations with long-range interactions (without hard truncation), it is necessary to subtract the $\vecq = \veczero$ component, which is equivalent to removing the mean value of the potential.
\begin{equation}
    \bar{W}^{\vecR}_{iijj} = {W}^{\vecR}_{iijj} - \frac{1}{N_{\vecR}} \sum_{\vecR}  {W}^{\vecR}_{iijj} ,
\end{equation}
where $N_{\vecR}$ is the number of cells in the lattice. Ultimately, the energy penalty from the long-range interaction cannot be avoided by the formation of localized stripes (which are computational artifacts due to the hard truncation), and the CDW state regains its stability.

\section{Concluding remarks}\label{sec:conclusion}

In this work we have constructed tight-binding based effective models for electron-doped 2D CrSBr. The hopping parameters were directly projected from the DFT Kohn-Sham Hamiltonian, and the Coulomb interactions were generated from constrained RPA calculations. The minimal model includes only one hopping matrix element within the quasi-1D chain, along with three Coulomb interaction parameters (intra-chain, nearest inter-chain, and second nearest inter-chain). For the full model, we incorporated all hopping parameters, cRPA Coulomb interactions, and a long-range part fitted with an anisotropic Rytova-Keldysh form. The phase diagrams were then calculated using periodic Hartree-Fock and density matrix embedding theory.

The electronic phase diagram was investigated by scanning the supercell size and varying initial electron densities. Several ground-state candidates were identified, including charge density wave (CDW) phases with different phase shifts, uniform state, inversion-symmetry broken state, and localized stripes. In the minimal model, the CDW state is found to be the ground state, consistent with Peierls' mechanism. The resulting CDW is well-fitted by a cosine wave, with its periodicity inversely proportional to the Fermi wave vector $\veck_{\text{F}}$ (or equivalently the doped electron concentration $n_{\text{elec}}$). The phase shift of the CDW depends on the competition between the nearest inter-chain and second nearest inter-chain interactions. When the nearest interaction dominates, the phase shift between chains is expected to be $\pi$, otherwise, the shift is $\pi / 2$.  

Going beyond the minimal model, we have considered additional Coulomb interaction parameters. The long-range Coulomb matrix elements were approximated by the real-space value of the screened Coulomb interaction, parametrized by generalized anisotropic Rytova-Keldysh potential. Evolving from short-range to long-range interactions, some meta-stable states (IB or localized stripes) can be computationally stabilized due to the boundary effect of truncation. These reflect the sensitivity of the phase diagram to the Coulomb interaction. Ultimately, when long-range interactions are fully considered, the CDW state is found to be the most stable phase.

Overall, the phases identified in the simulations are energetically very close 
(in the magnitude of several meV per chemical formula)
, so caution is necessary when interpreting the results of the low-energy model. There are several approximations to consider: (\lroman{1}) The model neglects lattice distortions (phonon degrees of freedom), and includes significant charge modulation (e.g. the large variation in site density of the CDW state), which naturally induces a tendency toward lattice distortion; (\lroman{2}) The model is based solely on the conduction bands, without considering valence modulation; (\lroman{3}) Doped electrons are introduced using the rigid band approximation, without explicitly modeling dopant atoms; (\lroman{4}) Only the density-density Coulomb interaction is included in the Hamiltonian. (\lroman{5}) The system can be viewed as a distorted (flat) triangular lattice. When sufficiently small doping is introduced, the system may form Wigner crystal states~\cite{Monarkha12-Wigner-crystal-review,Tsui24-wigner-crystal-expt}, where the charges condense at local regions.


All of these factors may substantially impact the detailed energetics of nearly degenerate states in doped materials, particularly when the system is excited or coupled to phonons/magnons~\cite{Bae22-exciton,Wang23-polariton,Ruta23-polariton-CrSBr,Diederich23-exciton-magnon-CrSBr,Jiang24-ultrafast-CrSBr,Datta24-CrSBr-magnon}, which would require further generalization of the current models.  
{The influence of phonons is expected to enhance the charge modulation via a pure Peierls distortion and thus enhance the CDW. Since the effective distortion potential caused by electron-phonon coupling is generally smaller than the electron-electron repulsion, the amplitude of enhancement is estimated to be small.}
While these factors may be important and should be investigated in the future, our work provided a baseline theoretical framework for such investigations and reveals a rich phase diagram for doped CrSBr. In particular, the balance between phase stability and properties, such as wavelength and phase shifts in the CDW phase, can be tuned by adjusting doping concentration or model parameters. This suggests a potential for fine-tuning quantum phases through synthetic manipulation of CrSBr and similar 2D materials.

\begin{acknowledgments}
	We thank Margalit Feuer, Michael Ziebel, Xavier Roy, Abhay Pasupathy, Timothy Berkelbach, Tianyu Zhu, and Shu Fay Ung for the helpful discussions. ZHC was funded by the
	Columbia MRSEC on Precision-Assembled Quantum Materials (PAQM) under award number
	DMR-2011738. The Flatiron
Institute is a division of the Simons Foundation.

Codes used in this work can be obtained from:
The \textsc{libDMET} repository is at
\url{https://github.com/gkclab/libdmet_preview}.
\textsc{PySCF} is available from \url{https://pyscf.org}.

\end{acknowledgments}



\bibliography{refs}

\begin{thebibliography}{65}%
\makeatletter
\providecommand \@ifxundefined [1]{%
 \@ifx{#1\undefined}
}%
\providecommand \@ifnum [1]{%
 \ifnum #1\expandafter \@firstoftwo
 \else \expandafter \@secondoftwo
 \fi
}%
\providecommand \@ifx [1]{%
 \ifx #1\expandafter \@firstoftwo
 \else \expandafter \@secondoftwo
 \fi
}%
\providecommand \natexlab [1]{#1}%
\providecommand \enquote  [1]{``#1''}%
\providecommand \bibnamefont  [1]{#1}%
\providecommand \bibfnamefont [1]{#1}%
\providecommand \citenamefont [1]{#1}%
\providecommand \href@noop [0]{\@secondoftwo}%
\providecommand \href [0]{\begingroup \@sanitize@url \@href}%
\providecommand \@href[1]{\@@startlink{#1}\@@href}%
\providecommand \@@href[1]{\endgroup#1\@@endlink}%
\providecommand \@sanitize@url [0]{\catcode `\\12\catcode `\$12\catcode
  `\&12\catcode `\#12\catcode `\^12\catcode `\_12\catcode `\%12\relax}%
\providecommand \@@startlink[1]{}%
\providecommand \@@endlink[0]{}%
\providecommand \url  [0]{\begingroup\@sanitize@url \@url }%
\providecommand \@url [1]{\endgroup\@href {#1}{\urlprefix }}%
\providecommand \urlprefix  [0]{URL }%
\providecommand \Eprint [0]{\href }%
\providecommand \doibase [0]{https://doi.org/}%
\providecommand \selectlanguage [0]{\@gobble}%
\providecommand \bibinfo  [0]{\@secondoftwo}%
\providecommand \bibfield  [0]{\@secondoftwo}%
\providecommand \translation [1]{[#1]}%
\providecommand \BibitemOpen [0]{}%
\providecommand \bibitemStop [0]{}%
\providecommand \bibitemNoStop [0]{.\EOS\space}%
\providecommand \EOS [0]{\spacefactor3000\relax}%
\providecommand \BibitemShut  [1]{\csname bibitem#1\endcsname}%
\let\auto@bib@innerbib\@empty
\bibitem [{\citenamefont {Ajayan}\ \emph {et~al.}(2016)\citenamefont {Ajayan},
  \citenamefont {Kim},\ and\ \citenamefont
  {Banerjee}}]{Ajayan16-2D-vdw-review}%
  \BibitemOpen
  \bibfield  {author} {\bibinfo {author} {\bibfnamefont {P.}~\bibnamefont
  {Ajayan}}, \bibinfo {author} {\bibfnamefont {P.}~\bibnamefont {Kim}},\ and\
  \bibinfo {author} {\bibfnamefont {K.}~\bibnamefont {Banerjee}},\ }\bibfield
  {title} {\bibinfo {title} {Two-dimensional {van der Waals} materials},\
  }\href@noop {} {\bibfield  {journal} {\bibinfo  {journal} {Phys. Today}\
  }\textbf {\bibinfo {volume} {69}},\ \bibinfo {pages} {38} (\bibinfo {year}
  {2016})}\BibitemShut {NoStop}%
\bibitem [{\citenamefont {Duong}\ \emph {et~al.}(2017)\citenamefont {Duong},
  \citenamefont {Yun},\ and\ \citenamefont {Lee}}]{Duong17-2D-vdw-review}%
  \BibitemOpen
  \bibfield  {author} {\bibinfo {author} {\bibfnamefont {D.~L.}\ \bibnamefont
  {Duong}}, \bibinfo {author} {\bibfnamefont {S.~J.}\ \bibnamefont {Yun}},\
  and\ \bibinfo {author} {\bibfnamefont {Y.~H.}\ \bibnamefont {Lee}},\
  }\bibfield  {title} {\bibinfo {title} {{van der Waals} layered materials:
  opportunities and challenges},\ }\href@noop {} {\bibfield  {journal}
  {\bibinfo  {journal} {ACS Nano}\ }\textbf {\bibinfo {volume} {11}},\ \bibinfo
  {pages} {11803} (\bibinfo {year} {2017})}\BibitemShut {NoStop}%
\bibitem [{\citenamefont {Burch}\ \emph {et~al.}(2018)\citenamefont {Burch},
  \citenamefont {Mandrus},\ and\ \citenamefont
  {Park}}]{Burch18-2D-vdw-magnetism}%
  \BibitemOpen
  \bibfield  {author} {\bibinfo {author} {\bibfnamefont {K.~S.}\ \bibnamefont
  {Burch}}, \bibinfo {author} {\bibfnamefont {D.}~\bibnamefont {Mandrus}},\
  and\ \bibinfo {author} {\bibfnamefont {J.-G.}\ \bibnamefont {Park}},\
  }\bibfield  {title} {\bibinfo {title} {Magnetism in two-dimensional {van der
  Waals} materials},\ }\href@noop {} {\bibfield  {journal} {\bibinfo  {journal}
  {Nature}\ }\textbf {\bibinfo {volume} {563}},\ \bibinfo {pages} {47}
  (\bibinfo {year} {2018})}\BibitemShut {NoStop}%
\bibitem [{\citenamefont {Wang}\ \emph {et~al.}(2022)\citenamefont {Wang},
  \citenamefont {Bedoya-Pinto}, \citenamefont {Blei}, \citenamefont {Dismukes},
  \citenamefont {Hamo}, \citenamefont {Jenkins}, \citenamefont {Koperski},
  \citenamefont {Liu}, \citenamefont {Sun}, \citenamefont {Telford},
  \citenamefont {Kim}, \citenamefont {Augustin}, \citenamefont {Vool},
  \citenamefont {Yin}, \citenamefont {Li}, \citenamefont {Falin}, \citenamefont
  {Dean}, \citenamefont {Casanova}, \citenamefont {Evans}, \citenamefont
  {Chshiev}, \citenamefont {Mishchenko}, \citenamefont {Petrovic},
  \citenamefont {He}, \citenamefont {Zhao}, \citenamefont {Tsen}, \citenamefont
  {Gerardot}, \citenamefont {Brotons-Gisbert}, \citenamefont {Guguchia},
  \citenamefont {Roy}, \citenamefont {Tongay}, \citenamefont {Wang},
  \citenamefont {Hasan}, \citenamefont {Wrachtrup}, \citenamefont {Yacoby},
  \citenamefont {Fert}, \citenamefont {Parkin}, \citenamefont {Novoselov},
  \citenamefont {Dai}, \citenamefont {Balicas},\ and\ \citenamefont
  {Santos}}]{Wang22-magnetic-2D-vdw-review}%
  \BibitemOpen
  \bibfield  {author} {\bibinfo {author} {\bibfnamefont {Q.~H.}\ \bibnamefont
  {Wang}}, \bibinfo {author} {\bibfnamefont {A.}~\bibnamefont {Bedoya-Pinto}},
  \bibinfo {author} {\bibfnamefont {M.}~\bibnamefont {Blei}}, \bibinfo {author}
  {\bibfnamefont {A.~H.}\ \bibnamefont {Dismukes}}, \bibinfo {author}
  {\bibfnamefont {A.}~\bibnamefont {Hamo}}, \bibinfo {author} {\bibfnamefont
  {S.}~\bibnamefont {Jenkins}}, \bibinfo {author} {\bibfnamefont
  {M.}~\bibnamefont {Koperski}}, \bibinfo {author} {\bibfnamefont
  {Y.}~\bibnamefont {Liu}}, \bibinfo {author} {\bibfnamefont {Q.-C.}\
  \bibnamefont {Sun}}, \bibinfo {author} {\bibfnamefont {E.~J.}\ \bibnamefont
  {Telford}}, \bibinfo {author} {\bibfnamefont {H.~H.}\ \bibnamefont {Kim}},
  \bibinfo {author} {\bibfnamefont {M.}~\bibnamefont {Augustin}}, \bibinfo
  {author} {\bibfnamefont {U.}~\bibnamefont {Vool}}, \bibinfo {author}
  {\bibfnamefont {J.-X.}\ \bibnamefont {Yin}}, \bibinfo {author} {\bibfnamefont
  {L.~H.}\ \bibnamefont {Li}}, \bibinfo {author} {\bibfnamefont
  {A.}~\bibnamefont {Falin}}, \bibinfo {author} {\bibfnamefont {C.~R.}\
  \bibnamefont {Dean}}, \bibinfo {author} {\bibfnamefont {F.}~\bibnamefont
  {Casanova}}, \bibinfo {author} {\bibfnamefont {R.~F.~L.}\ \bibnamefont
  {Evans}}, \bibinfo {author} {\bibfnamefont {M.}~\bibnamefont {Chshiev}},
  \bibinfo {author} {\bibfnamefont {A.}~\bibnamefont {Mishchenko}}, \bibinfo
  {author} {\bibfnamefont {C.}~\bibnamefont {Petrovic}}, \bibinfo {author}
  {\bibfnamefont {R.}~\bibnamefont {He}}, \bibinfo {author} {\bibfnamefont
  {L.}~\bibnamefont {Zhao}}, \bibinfo {author} {\bibfnamefont {A.~W.}\
  \bibnamefont {Tsen}}, \bibinfo {author} {\bibfnamefont {B.~D.}\ \bibnamefont
  {Gerardot}}, \bibinfo {author} {\bibfnamefont {M.}~\bibnamefont
  {Brotons-Gisbert}}, \bibinfo {author} {\bibfnamefont {Z.}~\bibnamefont
  {Guguchia}}, \bibinfo {author} {\bibfnamefont {X.}~\bibnamefont {Roy}},
  \bibinfo {author} {\bibfnamefont {S.}~\bibnamefont {Tongay}}, \bibinfo
  {author} {\bibfnamefont {Z.}~\bibnamefont {Wang}}, \bibinfo {author}
  {\bibfnamefont {M.~Z.}\ \bibnamefont {Hasan}}, \bibinfo {author}
  {\bibfnamefont {J.}~\bibnamefont {Wrachtrup}}, \bibinfo {author}
  {\bibfnamefont {A.}~\bibnamefont {Yacoby}}, \bibinfo {author} {\bibfnamefont
  {A.}~\bibnamefont {Fert}}, \bibinfo {author} {\bibfnamefont {S.}~\bibnamefont
  {Parkin}}, \bibinfo {author} {\bibfnamefont {K.~S.}\ \bibnamefont
  {Novoselov}}, \bibinfo {author} {\bibfnamefont {P.}~\bibnamefont {Dai}},
  \bibinfo {author} {\bibfnamefont {L.}~\bibnamefont {Balicas}},\ and\ \bibinfo
  {author} {\bibfnamefont {E.~J.~G.}\ \bibnamefont {Santos}},\ }\bibfield
  {title} {\bibinfo {title} {The magnetic genome of two-dimensional {van der
  Waals} materials},\ }\href {https://doi.org/10.1021/acsnano.1c09150}
  {\bibfield  {journal} {\bibinfo  {journal} {ACS Nano}\ }\textbf {\bibinfo
  {volume} {16}},\ \bibinfo {pages} {6960} (\bibinfo {year}
  {2022})}\BibitemShut {NoStop}%
\bibitem [{\citenamefont {Wilson}\ \emph {et~al.}(2021)\citenamefont {Wilson},
  \citenamefont {Lee}, \citenamefont {Cenker}, \citenamefont {Xie},
  \citenamefont {Dismukes}, \citenamefont {Telford}, \citenamefont {Fonseca},
  \citenamefont {Sivakumar}, \citenamefont {Dean}, \citenamefont {Cao},
  \citenamefont {Roy}, \citenamefont {Xu},\ and\ \citenamefont
  {Zhu}}]{Wilson21-CrSBr-interlayer}%
  \BibitemOpen
  \bibfield  {author} {\bibinfo {author} {\bibfnamefont {N.~P.}\ \bibnamefont
  {Wilson}}, \bibinfo {author} {\bibfnamefont {K.}~\bibnamefont {Lee}},
  \bibinfo {author} {\bibfnamefont {J.}~\bibnamefont {Cenker}}, \bibinfo
  {author} {\bibfnamefont {K.}~\bibnamefont {Xie}}, \bibinfo {author}
  {\bibfnamefont {A.~H.}\ \bibnamefont {Dismukes}}, \bibinfo {author}
  {\bibfnamefont {E.~J.}\ \bibnamefont {Telford}}, \bibinfo {author}
  {\bibfnamefont {J.}~\bibnamefont {Fonseca}}, \bibinfo {author} {\bibfnamefont
  {S.}~\bibnamefont {Sivakumar}}, \bibinfo {author} {\bibfnamefont
  {C.}~\bibnamefont {Dean}}, \bibinfo {author} {\bibfnamefont {T.}~\bibnamefont
  {Cao}}, \bibinfo {author} {\bibfnamefont {X.}~\bibnamefont {Roy}}, \bibinfo
  {author} {\bibfnamefont {X.}~\bibnamefont {Xu}},\ and\ \bibinfo {author}
  {\bibfnamefont {X.}~\bibnamefont {Zhu}},\ }\bibfield  {title} {\bibinfo
  {title} {Interlayer electronic coupling on demand in a {2D} magnetic
  semiconductor},\ }\href@noop {} {\bibfield  {journal} {\bibinfo  {journal}
  {Nat. Mater.}\ }\textbf {\bibinfo {volume} {20}},\ \bibinfo {pages} {1657}
  (\bibinfo {year} {2021})}\BibitemShut {NoStop}%
\bibitem [{\citenamefont {Cenker}\ \emph {et~al.}(2022)\citenamefont {Cenker},
  \citenamefont {Sivakumar}, \citenamefont {Xie}, \citenamefont {Miller},
  \citenamefont {Thijssen}, \citenamefont {Liu}, \citenamefont {Dismukes},
  \citenamefont {Fonseca}, \citenamefont {Anderson}, \citenamefont {Zhu},
  \citenamefont {Roy}, \citenamefont {Xiao}, \citenamefont {Chu}, \citenamefont
  {Cao},\ and\ \citenamefont {Xu}}]{Cenker22-CrSBr-strain}%
  \BibitemOpen
  \bibfield  {author} {\bibinfo {author} {\bibfnamefont {J.}~\bibnamefont
  {Cenker}}, \bibinfo {author} {\bibfnamefont {S.}~\bibnamefont {Sivakumar}},
  \bibinfo {author} {\bibfnamefont {K.}~\bibnamefont {Xie}}, \bibinfo {author}
  {\bibfnamefont {A.}~\bibnamefont {Miller}}, \bibinfo {author} {\bibfnamefont
  {P.}~\bibnamefont {Thijssen}}, \bibinfo {author} {\bibfnamefont
  {Z.}~\bibnamefont {Liu}}, \bibinfo {author} {\bibfnamefont {A.}~\bibnamefont
  {Dismukes}}, \bibinfo {author} {\bibfnamefont {J.}~\bibnamefont {Fonseca}},
  \bibinfo {author} {\bibfnamefont {E.}~\bibnamefont {Anderson}}, \bibinfo
  {author} {\bibfnamefont {X.}~\bibnamefont {Zhu}}, \bibinfo {author}
  {\bibfnamefont {X.}~\bibnamefont {Roy}}, \bibinfo {author} {\bibfnamefont
  {D.}~\bibnamefont {Xiao}}, \bibinfo {author} {\bibfnamefont {J.-H.}\
  \bibnamefont {Chu}}, \bibinfo {author} {\bibfnamefont {T.}~\bibnamefont
  {Cao}},\ and\ \bibinfo {author} {\bibfnamefont {X.}~\bibnamefont {Xu}},\
  }\bibfield  {title} {\bibinfo {title} {Reversible strain-induced magnetic
  phase transition in a {van der Waals} magnet},\ }\href@noop {} {\bibfield
  {journal} {\bibinfo  {journal} {Nat. Nanotechnol.}\ }\textbf {\bibinfo
  {volume} {17}},\ \bibinfo {pages} {256} (\bibinfo {year} {2022})}\BibitemShut
  {NoStop}%
\bibitem [{\citenamefont {Klein}\ \emph {et~al.}(2023)\citenamefont {Klein},
  \citenamefont {Pingault}, \citenamefont {Florian}, \citenamefont
  {Heißenb{\"u}ttel}, \citenamefont {Steinhoff}, \citenamefont {Song},
  \citenamefont {Torres}, \citenamefont {Dirnberger}, \citenamefont {Curtis},
  \citenamefont {Weile}, \citenamefont {Penn}, \citenamefont {Deilmann},
  \citenamefont {Dana}, \citenamefont {Bushati}, \citenamefont {Quan},
  \citenamefont {Luxa}, \citenamefont {Sofer}, \citenamefont {Alù},
  \citenamefont {Menon}, \citenamefont {Wurstbauer}, \citenamefont {Rohlfing},
  \citenamefont {Narang}, \citenamefont {Lončar},\ and\ \citenamefont
  {Ross}}]{Klein23-CrSBr-1D}%
  \BibitemOpen
  \bibfield  {author} {\bibinfo {author} {\bibfnamefont {J.}~\bibnamefont
  {Klein}}, \bibinfo {author} {\bibfnamefont {B.}~\bibnamefont {Pingault}},
  \bibinfo {author} {\bibfnamefont {M.}~\bibnamefont {Florian}}, \bibinfo
  {author} {\bibfnamefont {M.-C.}\ \bibnamefont {Heißenb{\"u}ttel}}, \bibinfo
  {author} {\bibfnamefont {A.}~\bibnamefont {Steinhoff}}, \bibinfo {author}
  {\bibfnamefont {Z.}~\bibnamefont {Song}}, \bibinfo {author} {\bibfnamefont
  {K.}~\bibnamefont {Torres}}, \bibinfo {author} {\bibfnamefont
  {F.}~\bibnamefont {Dirnberger}}, \bibinfo {author} {\bibfnamefont {J.~B.}\
  \bibnamefont {Curtis}}, \bibinfo {author} {\bibfnamefont {M.}~\bibnamefont
  {Weile}}, \bibinfo {author} {\bibfnamefont {A.}~\bibnamefont {Penn}},
  \bibinfo {author} {\bibfnamefont {T.}~\bibnamefont {Deilmann}}, \bibinfo
  {author} {\bibfnamefont {R.}~\bibnamefont {Dana}}, \bibinfo {author}
  {\bibfnamefont {R.}~\bibnamefont {Bushati}}, \bibinfo {author} {\bibfnamefont
  {J.}~\bibnamefont {Quan}}, \bibinfo {author} {\bibfnamefont {J.}~\bibnamefont
  {Luxa}}, \bibinfo {author} {\bibfnamefont {Z.}~\bibnamefont {Sofer}},
  \bibinfo {author} {\bibfnamefont {A.}~\bibnamefont {Alù}}, \bibinfo {author}
  {\bibfnamefont {V.~M.}\ \bibnamefont {Menon}}, \bibinfo {author}
  {\bibfnamefont {U.}~\bibnamefont {Wurstbauer}}, \bibinfo {author}
  {\bibfnamefont {M.}~\bibnamefont {Rohlfing}}, \bibinfo {author}
  {\bibfnamefont {P.}~\bibnamefont {Narang}}, \bibinfo {author} {\bibfnamefont
  {M.}~\bibnamefont {Lončar}},\ and\ \bibinfo {author} {\bibfnamefont {F.~M.}\
  \bibnamefont {Ross}},\ }\bibfield  {title} {\bibinfo {title} {The bulk {van
  der Waals} layered magnet {CrSBr} is a quasi-1d material},\ }\href
  {https://doi.org/10.1021/acsnano.2c07316} {\bibfield  {journal} {\bibinfo
  {journal} {ACS Nano}\ }\textbf {\bibinfo {volume} {17}},\ \bibinfo {pages}
  {5316} (\bibinfo {year} {2023})}\BibitemShut {NoStop}%
\bibitem [{\citenamefont {Long}\ \emph {et~al.}(2023)\citenamefont {Long},
  \citenamefont {Ghorbani-Asl}, \citenamefont {Mosina}, \citenamefont {Li},
  \citenamefont {Lin}, \citenamefont {Ganss}, \citenamefont {H{\"u}bner},
  \citenamefont {Sofer}, \citenamefont {Dirnberger}, \citenamefont {Kamra},
  \citenamefont {Krasheninnikov}, \citenamefont {Prucnal}, \citenamefont
  {Helm},\ and\ \citenamefont {Zhou}}]{Long23-CrSBr-ferromagnetic}%
  \BibitemOpen
  \bibfield  {author} {\bibinfo {author} {\bibfnamefont {F.}~\bibnamefont
  {Long}}, \bibinfo {author} {\bibfnamefont {M.}~\bibnamefont {Ghorbani-Asl}},
  \bibinfo {author} {\bibfnamefont {K.}~\bibnamefont {Mosina}}, \bibinfo
  {author} {\bibfnamefont {Y.}~\bibnamefont {Li}}, \bibinfo {author}
  {\bibfnamefont {K.}~\bibnamefont {Lin}}, \bibinfo {author} {\bibfnamefont
  {F.}~\bibnamefont {Ganss}}, \bibinfo {author} {\bibfnamefont
  {R.}~\bibnamefont {H{\"u}bner}}, \bibinfo {author} {\bibfnamefont
  {Z.}~\bibnamefont {Sofer}}, \bibinfo {author} {\bibfnamefont
  {F.}~\bibnamefont {Dirnberger}}, \bibinfo {author} {\bibfnamefont
  {A.}~\bibnamefont {Kamra}}, \bibinfo {author} {\bibfnamefont {A.~V.}\
  \bibnamefont {Krasheninnikov}}, \bibinfo {author} {\bibfnamefont
  {S.}~\bibnamefont {Prucnal}}, \bibinfo {author} {\bibfnamefont
  {M.}~\bibnamefont {Helm}},\ and\ \bibinfo {author} {\bibfnamefont
  {S.}~\bibnamefont {Zhou}},\ }\bibfield  {title} {\bibinfo {title}
  {Ferromagnetic interlayer coupling in {CrSBr} crystals irradiated by ions},\
  }\href {https://doi.org/10.1021/acs.nanolett.3c01920} {\bibfield  {journal}
  {\bibinfo  {journal} {Nano Lett.}\ }\textbf {\bibinfo {volume} {23}},\
  \bibinfo {pages} {8468} (\bibinfo {year} {2023})}\BibitemShut {NoStop}%
\bibitem [{\citenamefont {Xie}\ \emph {et~al.}(2023)\citenamefont {Xie},
  \citenamefont {Zhang}, \citenamefont {Xiao},\ and\ \citenamefont
  {Cao}}]{Xie23-CrSBr-AFM-FM}%
  \BibitemOpen
  \bibfield  {author} {\bibinfo {author} {\bibfnamefont {K.}~\bibnamefont
  {Xie}}, \bibinfo {author} {\bibfnamefont {X.-W.}\ \bibnamefont {Zhang}},
  \bibinfo {author} {\bibfnamefont {D.}~\bibnamefont {Xiao}},\ and\ \bibinfo
  {author} {\bibfnamefont {T.}~\bibnamefont {Cao}},\ }\bibfield  {title}
  {\bibinfo {title} {Engineering magnetic phases of layered antiferromagnets by
  interfacial charge transfer},\ }\href@noop {} {\bibfield  {journal} {\bibinfo
   {journal} {ACS nano}\ }\textbf {\bibinfo {volume} {17}},\ \bibinfo {pages}
  {22684} (\bibinfo {year} {2023})}\BibitemShut {NoStop}%
\bibitem [{\citenamefont {Dirnberger}\ \emph {et~al.}(2023)\citenamefont
  {Dirnberger}, \citenamefont {Quan}, \citenamefont {Bushati}, \citenamefont
  {Diederich}, \citenamefont {Florian}, \citenamefont {Klein}, \citenamefont
  {Mosina}, \citenamefont {Sofer}, \citenamefont {Xu}, \citenamefont {Kamra},
  \citenamefont {García-Vidal}, \citenamefont {Al\`u},\ and\ \citenamefont
  {Menon}}]{Dirnberger23-CrSBr-magnon-polariton}%
  \BibitemOpen
  \bibfield  {author} {\bibinfo {author} {\bibfnamefont {F.}~\bibnamefont
  {Dirnberger}}, \bibinfo {author} {\bibfnamefont {J.}~\bibnamefont {Quan}},
  \bibinfo {author} {\bibfnamefont {R.}~\bibnamefont {Bushati}}, \bibinfo
  {author} {\bibfnamefont {G.~M.}\ \bibnamefont {Diederich}}, \bibinfo {author}
  {\bibfnamefont {M.}~\bibnamefont {Florian}}, \bibinfo {author} {\bibfnamefont
  {J.}~\bibnamefont {Klein}}, \bibinfo {author} {\bibfnamefont
  {K.}~\bibnamefont {Mosina}}, \bibinfo {author} {\bibfnamefont
  {Z.}~\bibnamefont {Sofer}}, \bibinfo {author} {\bibfnamefont
  {X.}~\bibnamefont {Xu}}, \bibinfo {author} {\bibfnamefont {A.}~\bibnamefont
  {Kamra}}, \bibinfo {author} {\bibfnamefont {F.~J.}\ \bibnamefont
  {García-Vidal}}, \bibinfo {author} {\bibfnamefont {A.}~\bibnamefont
  {Al\`u}},\ and\ \bibinfo {author} {\bibfnamefont {V.~M.}\ \bibnamefont
  {Menon}},\ }\bibfield  {title} {\bibinfo {title} {Magneto-optics in a {van
  der Waals} magnet tuned by self-hybridized polaritons},\ }\href@noop {}
  {\bibfield  {journal} {\bibinfo  {journal} {Nature}\ }\textbf {\bibinfo
  {volume} {620}},\ \bibinfo {pages} {533} (\bibinfo {year}
  {2023})}\BibitemShut {NoStop}%
\bibitem [{\citenamefont {Feuer}\ \emph {et~al.}(2024)\citenamefont {Feuer},
  \citenamefont {Thinel}, \citenamefont {Huang}, \citenamefont {Cui},
  \citenamefont {Shao}, \citenamefont {Kundu}, \citenamefont {Chica},
  \citenamefont {Han}, \citenamefont {Pokratath}, \citenamefont {Telford},
  \citenamefont {Cox}, \citenamefont {York}, \citenamefont {Okuno},
  \citenamefont {Huang}, \citenamefont {Bukula}, \citenamefont {Nashabeh},
  \citenamefont {Qiu}, \citenamefont {Nuckolls}, \citenamefont {Dean},
  \citenamefont {Billinge}, \citenamefont {Zhu}, \citenamefont {Zhu},
  \citenamefont {Basov}, \citenamefont {Millis}, \citenamefont {Reichman},
  \citenamefont {Pasupathy}, \citenamefont {Roy},\ and\ \citenamefont
  {Ziebel}}]{Feuer24-CrSBr-doping}%
  \BibitemOpen
  \bibfield  {author} {\bibinfo {author} {\bibfnamefont {M.~L.}\ \bibnamefont
  {Feuer}}, \bibinfo {author} {\bibfnamefont {M.}~\bibnamefont {Thinel}},
  \bibinfo {author} {\bibfnamefont {X.}~\bibnamefont {Huang}}, \bibinfo
  {author} {\bibfnamefont {Z.-H.}\ \bibnamefont {Cui}}, \bibinfo {author}
  {\bibfnamefont {Y.}~\bibnamefont {Shao}}, \bibinfo {author} {\bibfnamefont
  {A.~K.}\ \bibnamefont {Kundu}}, \bibinfo {author} {\bibfnamefont {D.~G.}\
  \bibnamefont {Chica}}, \bibinfo {author} {\bibfnamefont {M.-G.}\ \bibnamefont
  {Han}}, \bibinfo {author} {\bibfnamefont {R.}~\bibnamefont {Pokratath}},
  \bibinfo {author} {\bibfnamefont {E.~J.}\ \bibnamefont {Telford}}, \bibinfo
  {author} {\bibfnamefont {J.}~\bibnamefont {Cox}}, \bibinfo {author}
  {\bibfnamefont {E.}~\bibnamefont {York}}, \bibinfo {author} {\bibfnamefont
  {S.}~\bibnamefont {Okuno}}, \bibinfo {author} {\bibfnamefont {C.-Y.}\
  \bibnamefont {Huang}}, \bibinfo {author} {\bibfnamefont {O.}~\bibnamefont
  {Bukula}}, \bibinfo {author} {\bibfnamefont {L.~M.}\ \bibnamefont
  {Nashabeh}}, \bibinfo {author} {\bibfnamefont {S.}~\bibnamefont {Qiu}},
  \bibinfo {author} {\bibfnamefont {C.~P.}\ \bibnamefont {Nuckolls}}, \bibinfo
  {author} {\bibfnamefont {C.~R.}\ \bibnamefont {Dean}}, \bibinfo {author}
  {\bibfnamefont {S.~J.~L.}\ \bibnamefont {Billinge}}, \bibinfo {author}
  {\bibfnamefont {X.}~\bibnamefont {Zhu}}, \bibinfo {author} {\bibfnamefont
  {Y.}~\bibnamefont {Zhu}}, \bibinfo {author} {\bibfnamefont {D.~N.}\
  \bibnamefont {Basov}}, \bibinfo {author} {\bibfnamefont {A.~J.}\ \bibnamefont
  {Millis}}, \bibinfo {author} {\bibfnamefont {D.~R.}\ \bibnamefont
  {Reichman}}, \bibinfo {author} {\bibfnamefont {A.~N.}\ \bibnamefont
  {Pasupathy}}, \bibinfo {author} {\bibfnamefont {X.}~\bibnamefont {Roy}},\
  and\ \bibinfo {author} {\bibfnamefont {M.~E.}\ \bibnamefont {Ziebel}},\
  }\bibfield  {title} {\bibinfo {title} {Doping-induced charge density wave and
  ferromagnetism in the van der {Waals} semiconductor {CrSBr}},\ }\href@noop {}
  {\bibfield  {journal} {\bibinfo  {journal} {arXiv preprint arXiv:2412.08631}\
  } (\bibinfo {year} {2024})}\BibitemShut {NoStop}%
\bibitem [{\citenamefont {Georges}\ \emph {et~al.}(1996)\citenamefont
  {Georges}, \citenamefont {Kotliar}, \citenamefont {Krauth},\ and\
  \citenamefont {Rozenberg}}]{Georges96}%
  \BibitemOpen
  \bibfield  {author} {\bibinfo {author} {\bibfnamefont {A.}~\bibnamefont
  {Georges}}, \bibinfo {author} {\bibfnamefont {G.}~\bibnamefont {Kotliar}},
  \bibinfo {author} {\bibfnamefont {W.}~\bibnamefont {Krauth}},\ and\ \bibinfo
  {author} {\bibfnamefont {M.~J.}\ \bibnamefont {Rozenberg}},\ }\bibfield
  {title} {\bibinfo {title} {Dynamical mean-field theory of strongly correlated
  fermion systems and the limit of infinite dimensions},\ }\href@noop {}
  {\bibfield  {journal} {\bibinfo  {journal} {Rev. Mod. Phys.}\ }\textbf
  {\bibinfo {volume} {68}},\ \bibinfo {pages} {13} (\bibinfo {year}
  {1996})}\BibitemShut {NoStop}%
\bibitem [{\citenamefont {Sun}\ and\ \citenamefont {Chan}(2016)}]{Sun16QET}%
  \BibitemOpen
  \bibfield  {author} {\bibinfo {author} {\bibfnamefont {Q.}~\bibnamefont
  {Sun}}\ and\ \bibinfo {author} {\bibfnamefont {G.~K.-L.}\ \bibnamefont
  {Chan}},\ }\bibfield  {title} {\bibinfo {title} {Quantum embedding
  theories},\ }\href@noop {} {\bibfield  {journal} {\bibinfo  {journal} {Acc.
  Chem. Res.}\ }\textbf {\bibinfo {volume} {49}},\ \bibinfo {pages} {2705}
  (\bibinfo {year} {2016})}\BibitemShut {NoStop}%
\bibitem [{\citenamefont {Gordon}(2017)}]{Gordon17-fragmentation-book}%
  \BibitemOpen
  \bibfield  {author} {\bibinfo {author} {\bibfnamefont {M.~S.}\ \bibnamefont
  {Gordon}},\ }\href@noop {} {\emph {\bibinfo {title} {Fragmentation: toward
  accurate calculations on complex molecular systems}}}\ (\bibinfo  {publisher}
  {John Wiley \& Sons},\ \bibinfo {year} {2017})\BibitemShut {NoStop}%
\bibitem [{\citenamefont {Ma}\ \emph {et~al.}(2021)\citenamefont {Ma},
  \citenamefont {Sheng}, \citenamefont {Govoni},\ and\ \citenamefont
  {Galli}}]{Ma21-quantum-impurity-embedding}%
  \BibitemOpen
  \bibfield  {author} {\bibinfo {author} {\bibfnamefont {H.}~\bibnamefont
  {Ma}}, \bibinfo {author} {\bibfnamefont {N.}~\bibnamefont {Sheng}}, \bibinfo
  {author} {\bibfnamefont {M.}~\bibnamefont {Govoni}},\ and\ \bibinfo {author}
  {\bibfnamefont {G.}~\bibnamefont {Galli}},\ }\bibfield  {title} {\bibinfo
  {title} {Quantum embedding theory for strongly correlated states in
  materials},\ }\href@noop {} {\bibfield  {journal} {\bibinfo  {journal} {{J.
  Chem. Theory Comput.}}\ }\textbf {\bibinfo {volume} {17}},\ \bibinfo {pages}
  {2116} (\bibinfo {year} {2021})}\BibitemShut {NoStop}%
\bibitem [{\citenamefont {Aryasetiawan}\ \emph {et~al.}(2004)\citenamefont
  {Aryasetiawan}, \citenamefont {Imada}, \citenamefont {Georges}, \citenamefont
  {Kotliar}, \citenamefont {Biermann},\ and\ \citenamefont
  {Lichtenstein}}]{Aryasetiawan04}%
  \BibitemOpen
  \bibfield  {author} {\bibinfo {author} {\bibfnamefont {F.}~\bibnamefont
  {Aryasetiawan}}, \bibinfo {author} {\bibfnamefont {M.}~\bibnamefont {Imada}},
  \bibinfo {author} {\bibfnamefont {A.}~\bibnamefont {Georges}}, \bibinfo
  {author} {\bibfnamefont {G.}~\bibnamefont {Kotliar}}, \bibinfo {author}
  {\bibfnamefont {S.}~\bibnamefont {Biermann}},\ and\ \bibinfo {author}
  {\bibfnamefont {A.~I.}\ \bibnamefont {Lichtenstein}},\ }\bibfield  {title}
  {\bibinfo {title} {{Frequency-dependent local interactions and low-energy
  effective models from electronic structure calculations}},\ }\href@noop {}
  {\bibfield  {journal} {\bibinfo  {journal} {Phys. Rev. B}\ }\textbf {\bibinfo
  {volume} {70}},\ \bibinfo {pages} {195104} (\bibinfo {year}
  {2004})}\BibitemShut {NoStop}%
\bibitem [{\citenamefont {Miyake}\ and\ \citenamefont
  {Aryasetiawan}(2008)}]{Miyake08}%
  \BibitemOpen
  \bibfield  {author} {\bibinfo {author} {\bibfnamefont {T.}~\bibnamefont
  {Miyake}}\ and\ \bibinfo {author} {\bibfnamefont {F.}~\bibnamefont
  {Aryasetiawan}},\ }\bibfield  {title} {\bibinfo {title} {{Screened Coulomb
  interaction in the maximally localized Wannier basis}},\ }\href@noop {}
  {\bibfield  {journal} {\bibinfo  {journal} {{Phys. Rev. B}}\ }\textbf
  {\bibinfo {volume} {77}},\ \bibinfo {pages} {085122} (\bibinfo {year}
  {2008})}\BibitemShut {NoStop}%
\bibitem [{\citenamefont {Vaugier}\ \emph {et~al.}(2012)\citenamefont
  {Vaugier}, \citenamefont {Jiang},\ and\ \citenamefont
  {Biermann}}]{Vaugier12}%
  \BibitemOpen
  \bibfield  {author} {\bibinfo {author} {\bibfnamefont {L.}~\bibnamefont
  {Vaugier}}, \bibinfo {author} {\bibfnamefont {H.}~\bibnamefont {Jiang}},\
  and\ \bibinfo {author} {\bibfnamefont {S.}~\bibnamefont {Biermann}},\
  }\bibfield  {title} {\bibinfo {title} {{Hubbard} {U} and {Hund} exchange {J}
  in transition metal oxides: Screening versus localization trends from
  constrained random phase approximation},\ }\href@noop {} {\bibfield
  {journal} {\bibinfo  {journal} {Phys. Rev. B}\ }\textbf {\bibinfo {volume}
  {86}},\ \bibinfo {pages} {165105} (\bibinfo {year} {2012})}\BibitemShut
  {NoStop}%
\bibitem [{\citenamefont {Knizia}\ and\ \citenamefont {Chan}(2012)}]{Knizia12}%
  \BibitemOpen
  \bibfield  {author} {\bibinfo {author} {\bibfnamefont {G.}~\bibnamefont
  {Knizia}}\ and\ \bibinfo {author} {\bibfnamefont {G.~K.-L.}\ \bibnamefont
  {Chan}},\ }\bibfield  {title} {\bibinfo {title} {Density matrix embedding: A
  simple alternative to dynamical mean-field theory},\ }\href@noop {}
  {\bibfield  {journal} {\bibinfo  {journal} {Phys. Rev. Lett.}\ }\textbf
  {\bibinfo {volume} {109}},\ \bibinfo {pages} {186404} (\bibinfo {year}
  {2012})}\BibitemShut {NoStop}%
\bibitem [{\citenamefont {Telford}\ \emph {et~al.}(2020)\citenamefont
  {Telford}, \citenamefont {Dismukes}, \citenamefont {Lee}, \citenamefont
  {Cheng}, \citenamefont {Wieteska}, \citenamefont {Bartholomew}, \citenamefont
  {Chen}, \citenamefont {Xu}, \citenamefont {Pasupathy}, \citenamefont {Zhu},\
  and\ \citenamefont {Dean}}]{Telford20-CrSBr}%
  \BibitemOpen
  \bibfield  {author} {\bibinfo {author} {\bibfnamefont {E.~J.}\ \bibnamefont
  {Telford}}, \bibinfo {author} {\bibfnamefont {A.~H.}\ \bibnamefont
  {Dismukes}}, \bibinfo {author} {\bibfnamefont {K.}~\bibnamefont {Lee}},
  \bibinfo {author} {\bibfnamefont {M.}~\bibnamefont {Cheng}}, \bibinfo
  {author} {\bibfnamefont {A.}~\bibnamefont {Wieteska}}, \bibinfo {author}
  {\bibfnamefont {A.~K.}\ \bibnamefont {Bartholomew}}, \bibinfo {author}
  {\bibfnamefont {Y.-S.}\ \bibnamefont {Chen}}, \bibinfo {author}
  {\bibfnamefont {X.}~\bibnamefont {Xu}}, \bibinfo {author} {\bibfnamefont
  {A.~N.}\ \bibnamefont {Pasupathy}}, \bibinfo {author} {\bibfnamefont
  {X.}~\bibnamefont {Zhu}},\ and\ \bibinfo {author} {\bibfnamefont {R.~X.}\
  \bibnamefont {Dean}, \bibfnamefont {Cory}},\ }\bibfield  {title} {\bibinfo
  {title} {Layered antiferromagnetism induces large negative magnetoresistance
  in the {van der Waals} semiconductor {CrSBr}},\ }\href@noop {} {\bibfield
  {journal} {\bibinfo  {journal} {Adv. Mater.}\ }\textbf {\bibinfo {volume}
  {32}},\ \bibinfo {pages} {2003240} (\bibinfo {year} {2020})}\BibitemShut
  {NoStop}%
\bibitem [{\citenamefont {Lee}\ \emph {et~al.}(2021)\citenamefont {Lee},
  \citenamefont {Dismukes}, \citenamefont {Telford}, \citenamefont {Wiscons},
  \citenamefont {Wang}, \citenamefont {Xu}, \citenamefont {Nuckolls},
  \citenamefont {Dean}, \citenamefont {Roy},\ and\ \citenamefont
  {Zhu}}]{Lee21-CrSBr-magnetic}%
  \BibitemOpen
  \bibfield  {author} {\bibinfo {author} {\bibfnamefont {K.}~\bibnamefont
  {Lee}}, \bibinfo {author} {\bibfnamefont {A.~H.}\ \bibnamefont {Dismukes}},
  \bibinfo {author} {\bibfnamefont {E.~J.}\ \bibnamefont {Telford}}, \bibinfo
  {author} {\bibfnamefont {R.~A.}\ \bibnamefont {Wiscons}}, \bibinfo {author}
  {\bibfnamefont {J.}~\bibnamefont {Wang}}, \bibinfo {author} {\bibfnamefont
  {X.}~\bibnamefont {Xu}}, \bibinfo {author} {\bibfnamefont {C.}~\bibnamefont
  {Nuckolls}}, \bibinfo {author} {\bibfnamefont {C.~R.}\ \bibnamefont {Dean}},
  \bibinfo {author} {\bibfnamefont {X.}~\bibnamefont {Roy}},\ and\ \bibinfo
  {author} {\bibfnamefont {X.}~\bibnamefont {Zhu}},\ }\bibfield  {title}
  {\bibinfo {title} {Magnetic order and symmetry in the {2D} semiconductor
  {CrSBr}},\ }\href@noop {} {\bibfield  {journal} {\bibinfo  {journal} {Nano
  Lett.}\ }\textbf {\bibinfo {volume} {21}},\ \bibinfo {pages} {3511} (\bibinfo
  {year} {2021})}\BibitemShut {NoStop}%
\bibitem [{\citenamefont {Rizzo}\ \emph {et~al.}(2022)\citenamefont {Rizzo},
  \citenamefont {McLeod}, \citenamefont {Carnahan}, \citenamefont {Telford},
  \citenamefont {Dismukes}, \citenamefont {Wiscons}, \citenamefont {Dong},
  \citenamefont {Nuckolls}, \citenamefont {Dean}, \citenamefont {Pasupathy},
  \citenamefont {Roy}, \citenamefont {Xiao},\ and\ \citenamefont
  {Basov}}]{Rizzo22-CrSBr-magnetism}%
  \BibitemOpen
  \bibfield  {author} {\bibinfo {author} {\bibfnamefont {D.~J.}\ \bibnamefont
  {Rizzo}}, \bibinfo {author} {\bibfnamefont {A.~S.}\ \bibnamefont {McLeod}},
  \bibinfo {author} {\bibfnamefont {C.}~\bibnamefont {Carnahan}}, \bibinfo
  {author} {\bibfnamefont {E.~J.}\ \bibnamefont {Telford}}, \bibinfo {author}
  {\bibfnamefont {A.~H.}\ \bibnamefont {Dismukes}}, \bibinfo {author}
  {\bibfnamefont {R.~A.}\ \bibnamefont {Wiscons}}, \bibinfo {author}
  {\bibfnamefont {Y.}~\bibnamefont {Dong}}, \bibinfo {author} {\bibfnamefont
  {C.}~\bibnamefont {Nuckolls}}, \bibinfo {author} {\bibfnamefont {C.~R.}\
  \bibnamefont {Dean}}, \bibinfo {author} {\bibfnamefont {A.~N.}\ \bibnamefont
  {Pasupathy}}, \bibinfo {author} {\bibfnamefont {X.}~\bibnamefont {Roy}},
  \bibinfo {author} {\bibfnamefont {D.}~\bibnamefont {Xiao}},\ and\ \bibinfo
  {author} {\bibfnamefont {D.~N.}\ \bibnamefont {Basov}},\ }\bibfield  {title}
  {\bibinfo {title} {Visualizing atomically layered magnetism in {CrSBr}},\
  }\href@noop {} {\bibfield  {journal} {\bibinfo  {journal} {Adv. Mater.}\
  }\textbf {\bibinfo {volume} {34}},\ \bibinfo {pages} {2201000} (\bibinfo
  {year} {2022})}\BibitemShut {NoStop}%
\bibitem [{\citenamefont {Bo}\ \emph {et~al.}(2023)\citenamefont {Bo},
  \citenamefont {Li}, \citenamefont {Xu}, \citenamefont {Wan},\ and\
  \citenamefont {Pu}}]{Bo23-CrSBr-magnetic}%
  \BibitemOpen
  \bibfield  {author} {\bibinfo {author} {\bibfnamefont {X.}~\bibnamefont
  {Bo}}, \bibinfo {author} {\bibfnamefont {F.}~\bibnamefont {Li}}, \bibinfo
  {author} {\bibfnamefont {X.}~\bibnamefont {Xu}}, \bibinfo {author}
  {\bibfnamefont {X.}~\bibnamefont {Wan}},\ and\ \bibinfo {author}
  {\bibfnamefont {Y.}~\bibnamefont {Pu}},\ }\bibfield  {title} {\bibinfo
  {title} {Calculated magnetic exchange interactions in the {van der Waals}
  layered magnet {CrSBr}},\ }\href@noop {} {\bibfield  {journal} {\bibinfo
  {journal} {New J. Phys.}\ }\textbf {\bibinfo {volume} {25}},\ \bibinfo
  {pages} {013026} (\bibinfo {year} {2023})}\BibitemShut {NoStop}%
\bibitem [{\citenamefont {Shen}\ \emph {et~al.}(2024)\citenamefont {Shen},
  \citenamefont {Xiao},\ and\ \citenamefont
  {Cao}}]{Shen24-exchange-interaction-CrSBr}%
  \BibitemOpen
  \bibfield  {author} {\bibinfo {author} {\bibfnamefont {L.}~\bibnamefont
  {Shen}}, \bibinfo {author} {\bibfnamefont {D.}~\bibnamefont {Xiao}},\ and\
  \bibinfo {author} {\bibfnamefont {T.}~\bibnamefont {Cao}},\ }\bibfield
  {title} {\bibinfo {title} {Proximity-induced exchange interaction: {A} new
  pathway for quantum sensing using spin centers in hexagonal boron nitride},\
  }\href@noop {} {\bibfield  {journal} {\bibinfo  {journal} {J. Phys. Chem.
  Lett.}\ }\textbf {\bibinfo {volume} {15}},\ \bibinfo {pages} {4359} (\bibinfo
  {year} {2024})}\BibitemShut {NoStop}%
\bibitem [{\citenamefont {Diederich}\ \emph {et~al.}(2024)\citenamefont
  {Diederich}, \citenamefont {Nguyen}, \citenamefont {Cenker}, \citenamefont
  {Fonseca}, \citenamefont {Pumulo}, \citenamefont {Bae}, \citenamefont
  {Chica}, \citenamefont {Roy}, \citenamefont {Zhu}, \citenamefont {Xiao},
  \citenamefont {Ren},\ and\ \citenamefont {Xu}}]{Diederich24-exciton-CrSBr}%
  \BibitemOpen
  \bibfield  {author} {\bibinfo {author} {\bibfnamefont {G.~M.}\ \bibnamefont
  {Diederich}}, \bibinfo {author} {\bibfnamefont {M.}~\bibnamefont {Nguyen}},
  \bibinfo {author} {\bibfnamefont {J.}~\bibnamefont {Cenker}}, \bibinfo
  {author} {\bibfnamefont {J.}~\bibnamefont {Fonseca}}, \bibinfo {author}
  {\bibfnamefont {S.}~\bibnamefont {Pumulo}}, \bibinfo {author} {\bibfnamefont
  {Y.~J.}\ \bibnamefont {Bae}}, \bibinfo {author} {\bibfnamefont {D.~G.}\
  \bibnamefont {Chica}}, \bibinfo {author} {\bibfnamefont {X.}~\bibnamefont
  {Roy}}, \bibinfo {author} {\bibfnamefont {X.}~\bibnamefont {Zhu}}, \bibinfo
  {author} {\bibfnamefont {D.}~\bibnamefont {Xiao}}, \bibinfo {author}
  {\bibfnamefont {Y.}~\bibnamefont {Ren}},\ and\ \bibinfo {author}
  {\bibfnamefont {X.}~\bibnamefont {Xu}},\ }\bibfield  {title} {\bibinfo
  {title} {Exciton dressing by extreme nonlinear magnons in a layered
  semiconductor},\ }\href@noop {} {\bibfield  {journal} {\bibinfo  {journal}
  {arXiv preprint arXiv:2411.14943}\ } (\bibinfo {year} {2024})}\BibitemShut
  {NoStop}%
\bibitem [{\citenamefont {Yang}\ \emph {et~al.}(2021)\citenamefont {Yang},
  \citenamefont {Wang}, \citenamefont {Liu}, \citenamefont {Lu},\ and\
  \citenamefont {Wu}}]{Yang21-CrSBr-magnetic}%
  \BibitemOpen
  \bibfield  {author} {\bibinfo {author} {\bibfnamefont {K.}~\bibnamefont
  {Yang}}, \bibinfo {author} {\bibfnamefont {G.}~\bibnamefont {Wang}}, \bibinfo
  {author} {\bibfnamefont {L.}~\bibnamefont {Liu}}, \bibinfo {author}
  {\bibfnamefont {D.}~\bibnamefont {Lu}},\ and\ \bibinfo {author}
  {\bibfnamefont {H.}~\bibnamefont {Wu}},\ }\bibfield  {title} {\bibinfo
  {title} {Triaxial magnetic anisotropy in the two-dimensional ferromagnetic
  semiconductor {CrSBr}},\ }\href@noop {} {\bibfield  {journal} {\bibinfo
  {journal} {{Phys. Rev. B}}\ }\textbf {\bibinfo {volume} {104}},\ \bibinfo
  {pages} {144416} (\bibinfo {year} {2021})}\BibitemShut {NoStop}%
\bibitem [{\citenamefont {Bianchi}\ \emph
  {et~al.}(2023{\natexlab{a}})\citenamefont {Bianchi}, \citenamefont {Acharya},
  \citenamefont {Dirnberger}, \citenamefont {Klein}, \citenamefont {Pashov},
  \citenamefont {Mosina}, \citenamefont {Sofer}, \citenamefont {Rudenko},
  \citenamefont {Katsnelson}, \citenamefont {van Schilfgaarde}, \citenamefont
  {R\"osner},\ and\ \citenamefont {Hofmann}}]{Bianchi23-CrSBr-gw-gap}%
  \BibitemOpen
  \bibfield  {author} {\bibinfo {author} {\bibfnamefont {M.}~\bibnamefont
  {Bianchi}}, \bibinfo {author} {\bibfnamefont {S.}~\bibnamefont {Acharya}},
  \bibinfo {author} {\bibfnamefont {F.}~\bibnamefont {Dirnberger}}, \bibinfo
  {author} {\bibfnamefont {J.}~\bibnamefont {Klein}}, \bibinfo {author}
  {\bibfnamefont {D.}~\bibnamefont {Pashov}}, \bibinfo {author} {\bibfnamefont
  {K.}~\bibnamefont {Mosina}}, \bibinfo {author} {\bibfnamefont
  {Z.}~\bibnamefont {Sofer}}, \bibinfo {author} {\bibfnamefont {A.~N.}\
  \bibnamefont {Rudenko}}, \bibinfo {author} {\bibfnamefont {M.~I.}\
  \bibnamefont {Katsnelson}}, \bibinfo {author} {\bibfnamefont
  {M.}~\bibnamefont {van Schilfgaarde}}, \bibinfo {author} {\bibfnamefont
  {M.}~\bibnamefont {R\"osner}},\ and\ \bibinfo {author} {\bibfnamefont
  {P.}~\bibnamefont {Hofmann}},\ }\bibfield  {title} {\bibinfo {title}
  {Paramagnetic electronic structure of {CrSBr}: Comparison between ab initio
  {$GW$} theory and angle-resolved photoemission spectroscopy},\ }\href
  {https://doi.org/10.1103/PhysRevB.107.235107} {\bibfield  {journal} {\bibinfo
   {journal} {{Phys. Rev. B}}\ }\textbf {\bibinfo {volume} {107}},\ \bibinfo
  {pages} {235107} (\bibinfo {year} {2023}{\natexlab{a}})}\BibitemShut
  {NoStop}%
\bibitem [{\citenamefont {Wouters}\ \emph {et~al.}(2016)\citenamefont
  {Wouters}, \citenamefont {Jim{\'e}nez-Hoyos}, \citenamefont {Sun},\ and\
  \citenamefont {Chan}}]{Wouters16}%
  \BibitemOpen
  \bibfield  {author} {\bibinfo {author} {\bibfnamefont {S.}~\bibnamefont
  {Wouters}}, \bibinfo {author} {\bibfnamefont {C.~A.}\ \bibnamefont
  {Jim{\'e}nez-Hoyos}}, \bibinfo {author} {\bibfnamefont {Q.}~\bibnamefont
  {Sun}},\ and\ \bibinfo {author} {\bibfnamefont {G.~K.-L.}\ \bibnamefont
  {Chan}},\ }\bibfield  {title} {\bibinfo {title} {A practical guide to density
  matrix embedding theory in quantum chemistry},\ }\href@noop {} {\bibfield
  {journal} {\bibinfo  {journal} {J. Chem. Theory Comput.}\ }\textbf {\bibinfo
  {volume} {12}},\ \bibinfo {pages} {2706} (\bibinfo {year}
  {2016})}\BibitemShut {NoStop}%
\bibitem [{\citenamefont {Cui}\ \emph {et~al.}(2020)\citenamefont {Cui},
  \citenamefont {Zhu},\ and\ \citenamefont {Chan}}]{Cui20-dmet-solid}%
  \BibitemOpen
  \bibfield  {author} {\bibinfo {author} {\bibfnamefont {Z.-H.}\ \bibnamefont
  {Cui}}, \bibinfo {author} {\bibfnamefont {T.}~\bibnamefont {Zhu}},\ and\
  \bibinfo {author} {\bibfnamefont {G.~K.-L.}\ \bibnamefont {Chan}},\
  }\bibfield  {title} {\bibinfo {title} {Efficient implementation of ab initio
  quantum embedding in periodic systems: Density matrix embedding theory},\
  }\href {https://pubs.acs.org/doi/full/10.1021/acs.jctc.9b00933} {\bibfield
  {journal} {\bibinfo  {journal} {J. Chem. Theory Comput.}\ }\textbf {\bibinfo
  {volume} {16}},\ \bibinfo {pages} {119} (\bibinfo {year} {2020})}\BibitemShut
  {NoStop}%
\bibitem [{\citenamefont {Cui}\ \emph {et~al.}(2022)\citenamefont {Cui},
  \citenamefont {Zhai}, \citenamefont {Zhang},\ and\ \citenamefont
  {Chan}}]{Cui21-cuprate-parent-state}%
  \BibitemOpen
  \bibfield  {author} {\bibinfo {author} {\bibfnamefont {Z.-H.}\ \bibnamefont
  {Cui}}, \bibinfo {author} {\bibfnamefont {H.}~\bibnamefont {Zhai}}, \bibinfo
  {author} {\bibfnamefont {X.}~\bibnamefont {Zhang}},\ and\ \bibinfo {author}
  {\bibfnamefont {G.~K.}\ \bibnamefont {Chan}},\ }\bibfield  {title} {\bibinfo
  {title} {Systematic electronic structure in the cuprate parent state from
  quantum many-body simulations},\ }\href
  {https://www.science.org/doi/10.1126/science.abm2295} {\bibfield  {journal}
  {\bibinfo  {journal} {Science}\ }\textbf {\bibinfo {volume} {377}},\ \bibinfo
  {pages} {1192} (\bibinfo {year} {2022})}\BibitemShut {NoStop}%
\bibitem [{\citenamefont {Shavitt}\ and\ \citenamefont
  {Bartlett}(2009)}]{Shavitt09book}%
  \BibitemOpen
  \bibfield  {author} {\bibinfo {author} {\bibfnamefont {I.}~\bibnamefont
  {Shavitt}}\ and\ \bibinfo {author} {\bibfnamefont {R.~J.}\ \bibnamefont
  {Bartlett}},\ }\href@noop {} {\emph {\bibinfo {title} {Many-Body methods in
  chemistry and physics: MBPT and coupled cluster Theory}}}\ (\bibinfo
  {publisher} {Cambridge University},\ \bibinfo {year} {2009})\BibitemShut
  {NoStop}%
\bibitem [{\citenamefont {L{\'o}pez-Paz}\ \emph {et~al.}(2022)\citenamefont
  {L{\'o}pez-Paz}, \citenamefont {Guguchia}, \citenamefont {Pomjakushin},
  \citenamefont {Witteveen}, \citenamefont {Cervellino}, \citenamefont
  {Luetkens}, \citenamefont {Casati}, \citenamefont {Morpurgo},\ and\
  \citenamefont {von Rohr}}]{Lopez22-CrSBr-struct}%
  \BibitemOpen
  \bibfield  {author} {\bibinfo {author} {\bibfnamefont {S.~A.}\ \bibnamefont
  {L{\'o}pez-Paz}}, \bibinfo {author} {\bibfnamefont {Z.}~\bibnamefont
  {Guguchia}}, \bibinfo {author} {\bibfnamefont {V.~Y.}\ \bibnamefont
  {Pomjakushin}}, \bibinfo {author} {\bibfnamefont {C.}~\bibnamefont
  {Witteveen}}, \bibinfo {author} {\bibfnamefont {A.}~\bibnamefont
  {Cervellino}}, \bibinfo {author} {\bibfnamefont {H.}~\bibnamefont
  {Luetkens}}, \bibinfo {author} {\bibfnamefont {N.}~\bibnamefont {Casati}},
  \bibinfo {author} {\bibfnamefont {A.~F.}\ \bibnamefont {Morpurgo}},\ and\
  \bibinfo {author} {\bibfnamefont {F.~O.}\ \bibnamefont {von Rohr}},\
  }\bibfield  {title} {\bibinfo {title} {Dynamic magnetic crossover at the
  origin of the hidden-order in {van der Waals} antiferromagnet {CrSBr}},\
  }\href@noop {} {\bibfield  {journal} {\bibinfo  {journal} {Nat. Commun.}\
  }\textbf {\bibinfo {volume} {13}},\ \bibinfo {pages} {4745} (\bibinfo {year}
  {2022})}\BibitemShut {NoStop}%
\bibitem [{\citenamefont {Perdew}\ \emph {et~al.}(1996)\citenamefont {Perdew},
  \citenamefont {Burke},\ and\ \citenamefont {Ernzerhof}}]{Perdew96PBE}%
  \BibitemOpen
  \bibfield  {author} {\bibinfo {author} {\bibfnamefont {J.~P.}\ \bibnamefont
  {Perdew}}, \bibinfo {author} {\bibfnamefont {K.}~\bibnamefont {Burke}},\ and\
  \bibinfo {author} {\bibfnamefont {M.}~\bibnamefont {Ernzerhof}},\ }\bibfield
  {title} {\bibinfo {title} {Generalized gradient approximation made simple},\
  }\href@noop {} {\bibfield  {journal} {\bibinfo  {journal} {{Phys. Rev.
  Lett.}}\ }\textbf {\bibinfo {volume} {77}},\ \bibinfo {pages} {3865}
  (\bibinfo {year} {1996})}\BibitemShut {NoStop}%
\bibitem [{\citenamefont {Liechtenstein}\ \emph {et~al.}(1995)\citenamefont
  {Liechtenstein}, \citenamefont {Anisimov},\ and\ \citenamefont
  {Zaanen}}]{Lichtenstein95}%
  \BibitemOpen
  \bibfield  {author} {\bibinfo {author} {\bibfnamefont {A.~I.}\ \bibnamefont
  {Liechtenstein}}, \bibinfo {author} {\bibfnamefont {V.~I.}\ \bibnamefont
  {Anisimov}},\ and\ \bibinfo {author} {\bibfnamefont {J.}~\bibnamefont
  {Zaanen}},\ }\bibfield  {title} {\bibinfo {title} {Density-functional theory
  and strong interactions: orbital ordering in {Mott}-{Hubbard} insulators},\
  }\href@noop {} {\bibfield  {journal} {\bibinfo  {journal} {{Phys. Rev. B}}\
  }\textbf {\bibinfo {volume} {52}},\ \bibinfo {pages} {R5467} (\bibinfo {year}
  {1995})}\BibitemShut {NoStop}%
\bibitem [{\citenamefont {Bianchi}\ \emph
  {et~al.}(2023{\natexlab{b}})\citenamefont {Bianchi}, \citenamefont {Hsieh},
  \citenamefont {Porat}, \citenamefont {Dirnberger}, \citenamefont {Klein},
  \citenamefont {Mosina}, \citenamefont {Sofer}, \citenamefont {Rudenko},
  \citenamefont {Katsnelson}, \citenamefont {Chen}, \citenamefont {R\"osner},\
  and\ \citenamefont {Hofmann}}]{Bianchi23-CrSBr-lifshitz-transition}%
  \BibitemOpen
  \bibfield  {author} {\bibinfo {author} {\bibfnamefont {M.}~\bibnamefont
  {Bianchi}}, \bibinfo {author} {\bibfnamefont {K.}~\bibnamefont {Hsieh}},
  \bibinfo {author} {\bibfnamefont {E.~J.}\ \bibnamefont {Porat}}, \bibinfo
  {author} {\bibfnamefont {F.}~\bibnamefont {Dirnberger}}, \bibinfo {author}
  {\bibfnamefont {J.}~\bibnamefont {Klein}}, \bibinfo {author} {\bibfnamefont
  {K.}~\bibnamefont {Mosina}}, \bibinfo {author} {\bibfnamefont
  {Z.}~\bibnamefont {Sofer}}, \bibinfo {author} {\bibfnamefont {A.~N.}\
  \bibnamefont {Rudenko}}, \bibinfo {author} {\bibfnamefont {M.~I.}\
  \bibnamefont {Katsnelson}}, \bibinfo {author} {\bibfnamefont {Y.~P.}\
  \bibnamefont {Chen}}, \bibinfo {author} {\bibfnamefont {M.}~\bibnamefont
  {R\"osner}},\ and\ \bibinfo {author} {\bibfnamefont {P.}~\bibnamefont
  {Hofmann}},\ }\bibfield  {title} {\bibinfo {title} {Charge transfer induced
  lifshitz transition and magnetic symmetry breaking in ultrathin {CrSBr}
  crystals},\ }\href {https://doi.org/10.1103/PhysRevB.108.195410} {\bibfield
  {journal} {\bibinfo  {journal} {{Phys. Rev. B}}\ }\textbf {\bibinfo {volume}
  {108}},\ \bibinfo {pages} {195410} (\bibinfo {year}
  {2023}{\natexlab{b}})}\BibitemShut {NoStop}%
\bibitem [{\citenamefont {Rudenko}\ \emph {et~al.}(2023)\citenamefont
  {Rudenko}, \citenamefont {R{\"o}sner},\ and\ \citenamefont
  {Katsnelson}}]{Rudenko23-CrSBr-dielectric}%
  \BibitemOpen
  \bibfield  {author} {\bibinfo {author} {\bibfnamefont {A.~N.}\ \bibnamefont
  {Rudenko}}, \bibinfo {author} {\bibfnamefont {M.}~\bibnamefont
  {R{\"o}sner}},\ and\ \bibinfo {author} {\bibfnamefont {M.~I.}\ \bibnamefont
  {Katsnelson}},\ }\bibfield  {title} {\bibinfo {title} {Dielectric tunability
  of magnetic properties in orthorhombic ferromagnetic monolayer {CrSBr}},\
  }\href@noop {} {\bibfield  {journal} {\bibinfo  {journal} {npj Comput.
  Mater.}\ }\textbf {\bibinfo {volume} {9}},\ \bibinfo {pages} {83} (\bibinfo
  {year} {2023})}\BibitemShut {NoStop}%
\bibitem [{\citenamefont {Kresse}\ and\ \citenamefont
  {Hafner}(1994)}]{Kresse94-vasp}%
  \BibitemOpen
  \bibfield  {author} {\bibinfo {author} {\bibfnamefont {G.}~\bibnamefont
  {Kresse}}\ and\ \bibinfo {author} {\bibfnamefont {J.}~\bibnamefont
  {Hafner}},\ }\bibfield  {title} {\bibinfo {title} {Ab initio
  molecular-dynamics simulation of the liquid-metal--amorphous-semiconductor
  transition in germanium},\ }\href@noop {} {\bibfield  {journal} {\bibinfo
  {journal} {{Phys. Rev. B}}\ }\textbf {\bibinfo {volume} {49}},\ \bibinfo
  {pages} {14251} (\bibinfo {year} {1994})}\BibitemShut {NoStop}%
\bibitem [{\citenamefont {Kresse}\ and\ \citenamefont
  {Furthm{\"u}ller}(1996)}]{Kresse96-vasp}%
  \BibitemOpen
  \bibfield  {author} {\bibinfo {author} {\bibfnamefont {G.}~\bibnamefont
  {Kresse}}\ and\ \bibinfo {author} {\bibfnamefont {J.}~\bibnamefont
  {Furthm{\"u}ller}},\ }\bibfield  {title} {\bibinfo {title} {Efficiency of
  ab-initio total energy calculations for metals and semiconductors using a
  plane-wave basis set},\ }\href@noop {} {\bibfield  {journal} {\bibinfo
  {journal} {Comp. Mater. Sci.}\ }\textbf {\bibinfo {volume} {6}},\ \bibinfo
  {pages} {15} (\bibinfo {year} {1996})}\BibitemShut {NoStop}%
\bibitem [{\citenamefont {Kresse}\ and\ \citenamefont
  {Furthm\"uller}(1996)}]{Kresse96}%
  \BibitemOpen
  \bibfield  {author} {\bibinfo {author} {\bibfnamefont {G.}~\bibnamefont
  {Kresse}}\ and\ \bibinfo {author} {\bibfnamefont {J.}~\bibnamefont
  {Furthm\"uller}},\ }\bibfield  {title} {\bibinfo {title} {Efficient iterative
  schemes for \textit{ab initio} total-energy calculations using a plane-wave
  basis set},\ }\href@noop {} {\bibfield  {journal} {\bibinfo  {journal} {Phys.
  Rev. B}\ }\textbf {\bibinfo {volume} {54}},\ \bibinfo {pages} {11169}
  (\bibinfo {year} {1996})}\BibitemShut {NoStop}%
\bibitem [{\citenamefont {Marzari}\ \emph {et~al.}(2012)\citenamefont
  {Marzari}, \citenamefont {Mostofi}, \citenamefont {Yates}, \citenamefont
  {Souza},\ and\ \citenamefont {Vanderbilt}}]{Marzari12RMP}%
  \BibitemOpen
  \bibfield  {author} {\bibinfo {author} {\bibfnamefont {N.}~\bibnamefont
  {Marzari}}, \bibinfo {author} {\bibfnamefont {A.~A.}\ \bibnamefont
  {Mostofi}}, \bibinfo {author} {\bibfnamefont {Y.~R.}\ \bibnamefont {Yates}},
  \bibinfo {author} {\bibfnamefont {I.}~\bibnamefont {Souza}},\ and\ \bibinfo
  {author} {\bibfnamefont {D.}~\bibnamefont {Vanderbilt}},\ }\bibfield  {title}
  {\bibinfo {title} {Maximally localized wannier functions: Theory and
  applications},\ }\href@noop {} {\bibfield  {journal} {\bibinfo  {journal}
  {Rev. Mod. Phys.}\ }\textbf {\bibinfo {volume} {84}},\ \bibinfo {pages} {1419
  } (\bibinfo {year} {2012})}\BibitemShut {NoStop}%
\bibitem [{\citenamefont {Mostofi}\ \emph {et~al.}(2014)\citenamefont
  {Mostofi}, \citenamefont {Yates}, \citenamefont {Pizzi}, \citenamefont {Lee},
  \citenamefont {Souza}, \citenamefont {Vanderbilt},\ and\ \citenamefont
  {Marzari}}]{Mostofi14}%
  \BibitemOpen
  \bibfield  {author} {\bibinfo {author} {\bibfnamefont {A.~A.}\ \bibnamefont
  {Mostofi}}, \bibinfo {author} {\bibfnamefont {J.~R.}\ \bibnamefont {Yates}},
  \bibinfo {author} {\bibfnamefont {G.}~\bibnamefont {Pizzi}}, \bibinfo
  {author} {\bibfnamefont {Y.-S.}\ \bibnamefont {Lee}}, \bibinfo {author}
  {\bibfnamefont {I.}~\bibnamefont {Souza}}, \bibinfo {author} {\bibfnamefont
  {D.}~\bibnamefont {Vanderbilt}},\ and\ \bibinfo {author} {\bibfnamefont
  {N.}~\bibnamefont {Marzari}},\ }\bibfield  {title} {\bibinfo {title} {An
  updated version of wannier90: {A} tool for obtaining maximally-localised
  {Wannier} functions},\ }\href@noop {} {\bibfield  {journal} {\bibinfo
  {journal} {Comput. Phys. Commun.}\ }\textbf {\bibinfo {volume} {185}},\
  \bibinfo {pages} {2309} (\bibinfo {year} {2014})}\BibitemShut {NoStop}%
\bibitem [{\citenamefont {Kaltak}(2015)}]{Kaltak15-cRPA-thesis}%
  \BibitemOpen
  \bibfield  {author} {\bibinfo {author} {\bibfnamefont {M.}~\bibnamefont
  {Kaltak}},\ }\emph {\bibinfo {title} {Merging GW with DMFT}},\ \href@noop {}
  {Ph.D. thesis},\ \bibinfo  {school} {University of Vienna Vienna, Austria}
  (\bibinfo {year} {2015})\BibitemShut {NoStop}%
\bibitem [{\citenamefont {Hedin}(1965)}]{Hedin65}%
  \BibitemOpen
  \bibfield  {author} {\bibinfo {author} {\bibfnamefont {L.}~\bibnamefont
  {Hedin}},\ }\bibfield  {title} {\bibinfo {title} {New method for calculating
  the one-particle green's function with application to the electron-gas
  problem},\ }\href@noop {} {\bibfield  {journal} {\bibinfo  {journal} {{Phys.
  Rev.}}\ }\textbf {\bibinfo {volume} {139}},\ \bibinfo {pages} {A796}
  (\bibinfo {year} {1965})}\BibitemShut {NoStop}%
\bibitem [{\citenamefont {Hybertsen}\ and\ \citenamefont
  {Louie}(1986)}]{Hybertsen86}%
  \BibitemOpen
  \bibfield  {author} {\bibinfo {author} {\bibfnamefont {M.~S.}\ \bibnamefont
  {Hybertsen}}\ and\ \bibinfo {author} {\bibfnamefont {S.~G.}\ \bibnamefont
  {Louie}},\ }\bibfield  {title} {\bibinfo {title} {{Electron correlation in
  semiconductors and insulators: Band gap and quasiparticle energies}},\
  }\href@noop {} {\bibfield  {journal} {\bibinfo  {journal} {{Phys. Rev. B}}\
  }\textbf {\bibinfo {volume} {34}},\ \bibinfo {pages} {5390} (\bibinfo {year}
  {1986})}\BibitemShut {NoStop}%
\bibitem [{\citenamefont {Aryasetiawan}\ and\ \citenamefont
  {Gunnarsson}(1998)}]{Aryasetiawan98RPP}%
  \BibitemOpen
  \bibfield  {author} {\bibinfo {author} {\bibfnamefont {F.}~\bibnamefont
  {Aryasetiawan}}\ and\ \bibinfo {author} {\bibfnamefont {O.}~\bibnamefont
  {Gunnarsson}},\ }\bibfield  {title} {\bibinfo {title} {{The $GW$ Method}},\
  }\href@noop {} {\bibfield  {journal} {\bibinfo  {journal} {Rep.\ Prog.\
  Phys.}\ }\textbf {\bibinfo {volume} {61}},\ \bibinfo {pages} {237} (\bibinfo
  {year} {1998})}\BibitemShut {NoStop}%
\bibitem [{\citenamefont {Shishkin}\ and\ \citenamefont
  {Kresse}(2006)}]{Shishkin06}%
  \BibitemOpen
  \bibfield  {author} {\bibinfo {author} {\bibfnamefont {M.}~\bibnamefont
  {Shishkin}}\ and\ \bibinfo {author} {\bibfnamefont {G.}~\bibnamefont
  {Kresse}},\ }\bibfield  {title} {\bibinfo {title} {{Implementation and
  Performance of the Frequency-dependent GW Method with the PAW Framework}},\
  }\href@noop {} {\bibfield  {journal} {\bibinfo  {journal} {{Phys. Rev. B}}\
  }\textbf {\bibinfo {volume} {74}},\ \bibinfo {pages} {035101} (\bibinfo
  {year} {2006})}\BibitemShut {NoStop}%
\bibitem [{\citenamefont {Cui}\ \emph {et~al.}(2025)\citenamefont {Cui},
  \citenamefont {Yang}, \citenamefont {T{\"o}lle}, \citenamefont {Ye},
  \citenamefont {Yuan}, \citenamefont {Zhai}, \citenamefont {Park},
  \citenamefont {Kim}, \citenamefont {Zhang}, \citenamefont {Lin},
  \citenamefont {Berkelbach},\ and\ \citenamefont
  {Chan}}]{Cui23-cuprate-doping}%
  \BibitemOpen
  \bibfield  {author} {\bibinfo {author} {\bibfnamefont {Z.-H.}\ \bibnamefont
  {Cui}}, \bibinfo {author} {\bibfnamefont {J.}~\bibnamefont {Yang}}, \bibinfo
  {author} {\bibfnamefont {J.}~\bibnamefont {T{\"o}lle}}, \bibinfo {author}
  {\bibfnamefont {H.-Z.}\ \bibnamefont {Ye}}, \bibinfo {author} {\bibfnamefont
  {S.}~\bibnamefont {Yuan}}, \bibinfo {author} {\bibfnamefont {H.}~\bibnamefont
  {Zhai}}, \bibinfo {author} {\bibfnamefont {G.}~\bibnamefont {Park}}, \bibinfo
  {author} {\bibfnamefont {R.}~\bibnamefont {Kim}}, \bibinfo {author}
  {\bibfnamefont {X.}~\bibnamefont {Zhang}}, \bibinfo {author} {\bibfnamefont
  {L.}~\bibnamefont {Lin}}, \bibinfo {author} {\bibfnamefont {T.~C.}\
  \bibnamefont {Berkelbach}},\ and\ \bibinfo {author} {\bibfnamefont
  {G.~K.-L.}\ \bibnamefont {Chan}},\ }\bibfield  {title} {\bibinfo {title} {Ab
  initio quantum many-body description of superconducting trends in the
  cuprates},\ }\href@noop {} {\bibfield  {journal} {\bibinfo  {journal} {Nat.
  Commun.}\ }\textbf {\bibinfo {volume} {16}},\ \bibinfo {pages} {1845}
  (\bibinfo {year} {2025})}\BibitemShut {NoStop}%
\bibitem [{\citenamefont {Sun}\ \emph {et~al.}(2018)\citenamefont {Sun},
  \citenamefont {Berkelbach}, \citenamefont {Blunt}, \citenamefont {Booth},
  \citenamefont {Guo}, \citenamefont {Li}, \citenamefont {Liu}, \citenamefont
  {McClain}, \citenamefont {Sayfutyarova}, \citenamefont {Sharma},
  \citenamefont {Wouters},\ and\ \citenamefont {Chan}}]{Sun18pyscf}%
  \BibitemOpen
  \bibfield  {author} {\bibinfo {author} {\bibfnamefont {Q.}~\bibnamefont
  {Sun}}, \bibinfo {author} {\bibfnamefont {T.~C.}\ \bibnamefont {Berkelbach}},
  \bibinfo {author} {\bibfnamefont {N.~S.}\ \bibnamefont {Blunt}}, \bibinfo
  {author} {\bibfnamefont {G.~H.}\ \bibnamefont {Booth}}, \bibinfo {author}
  {\bibfnamefont {S.}~\bibnamefont {Guo}}, \bibinfo {author} {\bibfnamefont
  {Z.}~\bibnamefont {Li}}, \bibinfo {author} {\bibfnamefont {J.}~\bibnamefont
  {Liu}}, \bibinfo {author} {\bibfnamefont {J.~D.}\ \bibnamefont {McClain}},
  \bibinfo {author} {\bibfnamefont {E.~R.}\ \bibnamefont {Sayfutyarova}},
  \bibinfo {author} {\bibfnamefont {S.}~\bibnamefont {Sharma}}, \bibinfo
  {author} {\bibfnamefont {S.}~\bibnamefont {Wouters}},\ and\ \bibinfo {author}
  {\bibfnamefont {G.~K.-L.}\ \bibnamefont {Chan}},\ }\bibfield  {title}
  {\bibinfo {title} {{PySCF}: the {Python}-based simulations of chemistry
  framework},\ }\href@noop {} {\bibfield  {journal} {\bibinfo  {journal} {WIREs
  Comput. Mol. Sci.}\ }\textbf {\bibinfo {volume} {8}},\ \bibinfo {pages}
  {e1340} (\bibinfo {year} {2018})}\BibitemShut {NoStop}%
\bibitem [{\citenamefont {Sun}\ \emph {et~al.}(2020)\citenamefont {Sun},
  \citenamefont {Zhang}, \citenamefont {Banerjee}, \citenamefont {Bao},
  \citenamefont {Barbry}, \citenamefont {Blunt}, \citenamefont {Bogdanov},
  \citenamefont {Booth}, \citenamefont {Chen}, \citenamefont {Cui},
  \citenamefont {Eriksen}, \citenamefont {Gao}, \citenamefont {Guo},
  \citenamefont {Hermann}, \citenamefont {Hermes}, \citenamefont {Koh},
  \citenamefont {Koval}, \citenamefont {Lehtola}, \citenamefont {Li},
  \citenamefont {Liu}, \citenamefont {Mardirossian}, \citenamefont {McClain},
  \citenamefont {Motta}, \citenamefont {Mussard}, \citenamefont {Pham},
  \citenamefont {Pulkin}, \citenamefont {Purwanto}, \citenamefont {Robinson},
  \citenamefont {Ronca}, \citenamefont {Sayfutyarova}, \citenamefont
  {Scheurer}, \citenamefont {Schurkus}, \citenamefont {Smith}, \citenamefont
  {Sun}, \citenamefont {Sun}, \citenamefont {Upadhyay}, \citenamefont {Wagner},
  \citenamefont {Wang}, \citenamefont {White}, \citenamefont {Whitfield},
  \citenamefont {Williamson}, \citenamefont {Wouters}, \citenamefont {Yang},
  \citenamefont {Yu}, \citenamefont {Zhu}, \citenamefont {Berkelbach},
  \citenamefont {Sharma}, \citenamefont {Sokolov},\ and\ \citenamefont
  {Chan}}]{Sun20pyscf}%
  \BibitemOpen
  \bibfield  {author} {\bibinfo {author} {\bibfnamefont {Q.}~\bibnamefont
  {Sun}}, \bibinfo {author} {\bibfnamefont {X.}~\bibnamefont {Zhang}}, \bibinfo
  {author} {\bibfnamefont {S.}~\bibnamefont {Banerjee}}, \bibinfo {author}
  {\bibfnamefont {P.}~\bibnamefont {Bao}}, \bibinfo {author} {\bibfnamefont
  {M.}~\bibnamefont {Barbry}}, \bibinfo {author} {\bibfnamefont {N.~S.}\
  \bibnamefont {Blunt}}, \bibinfo {author} {\bibfnamefont {N.~A.}\ \bibnamefont
  {Bogdanov}}, \bibinfo {author} {\bibfnamefont {G.~H.}\ \bibnamefont {Booth}},
  \bibinfo {author} {\bibfnamefont {J.}~\bibnamefont {Chen}}, \bibinfo {author}
  {\bibfnamefont {Z.-H.}\ \bibnamefont {Cui}}, \bibinfo {author} {\bibfnamefont
  {J.~J.}\ \bibnamefont {Eriksen}}, \bibinfo {author} {\bibfnamefont
  {Y.}~\bibnamefont {Gao}}, \bibinfo {author} {\bibfnamefont {S.}~\bibnamefont
  {Guo}}, \bibinfo {author} {\bibfnamefont {J.}~\bibnamefont {Hermann}},
  \bibinfo {author} {\bibfnamefont {M.~R.}\ \bibnamefont {Hermes}}, \bibinfo
  {author} {\bibfnamefont {K.}~\bibnamefont {Koh}}, \bibinfo {author}
  {\bibfnamefont {P.}~\bibnamefont {Koval}}, \bibinfo {author} {\bibfnamefont
  {S.}~\bibnamefont {Lehtola}}, \bibinfo {author} {\bibfnamefont
  {Z.}~\bibnamefont {Li}}, \bibinfo {author} {\bibfnamefont {J.}~\bibnamefont
  {Liu}}, \bibinfo {author} {\bibfnamefont {N.}~\bibnamefont {Mardirossian}},
  \bibinfo {author} {\bibfnamefont {J.~D.}\ \bibnamefont {McClain}}, \bibinfo
  {author} {\bibfnamefont {M.}~\bibnamefont {Motta}}, \bibinfo {author}
  {\bibfnamefont {B.}~\bibnamefont {Mussard}}, \bibinfo {author} {\bibfnamefont
  {H.~Q.}\ \bibnamefont {Pham}}, \bibinfo {author} {\bibfnamefont
  {A.}~\bibnamefont {Pulkin}}, \bibinfo {author} {\bibfnamefont
  {W.}~\bibnamefont {Purwanto}}, \bibinfo {author} {\bibfnamefont {P.~J.}\
  \bibnamefont {Robinson}}, \bibinfo {author} {\bibfnamefont {E.}~\bibnamefont
  {Ronca}}, \bibinfo {author} {\bibfnamefont {E.}~\bibnamefont {Sayfutyarova}},
  \bibinfo {author} {\bibfnamefont {M.}~\bibnamefont {Scheurer}}, \bibinfo
  {author} {\bibfnamefont {H.~F.}\ \bibnamefont {Schurkus}}, \bibinfo {author}
  {\bibfnamefont {J.~E.~T.}\ \bibnamefont {Smith}}, \bibinfo {author}
  {\bibfnamefont {C.}~\bibnamefont {Sun}}, \bibinfo {author} {\bibfnamefont
  {S.-N.}\ \bibnamefont {Sun}}, \bibinfo {author} {\bibfnamefont
  {S.}~\bibnamefont {Upadhyay}}, \bibinfo {author} {\bibfnamefont {L.~K.}\
  \bibnamefont {Wagner}}, \bibinfo {author} {\bibfnamefont {X.}~\bibnamefont
  {Wang}}, \bibinfo {author} {\bibfnamefont {A.}~\bibnamefont {White}},
  \bibinfo {author} {\bibfnamefont {J.~D.}\ \bibnamefont {Whitfield}}, \bibinfo
  {author} {\bibfnamefont {M.~J.}\ \bibnamefont {Williamson}}, \bibinfo
  {author} {\bibfnamefont {S.}~\bibnamefont {Wouters}}, \bibinfo {author}
  {\bibfnamefont {J.}~\bibnamefont {Yang}}, \bibinfo {author} {\bibfnamefont
  {J.~M.}\ \bibnamefont {Yu}}, \bibinfo {author} {\bibfnamefont
  {T.}~\bibnamefont {Zhu}}, \bibinfo {author} {\bibfnamefont {T.~C.}\
  \bibnamefont {Berkelbach}}, \bibinfo {author} {\bibfnamefont
  {S.}~\bibnamefont {Sharma}}, \bibinfo {author} {\bibfnamefont
  {A.}~\bibnamefont {Sokolov}},\ and\ \bibinfo {author} {\bibfnamefont
  {G.~K.-L.}\ \bibnamefont {Chan}},\ }\bibfield  {title} {\bibinfo {title}
  {Recent developments in the {PySCF} program package},\ }\href
  {https://aip.scitation.org/doi/full/10.1063/5.0006074} {\bibfield  {journal}
  {\bibinfo  {journal} {J. Chem. Phys.}\ }\textbf {\bibinfo {volume} {153}},\
  \bibinfo {pages} {024109} (\bibinfo {year} {2020})}\BibitemShut {NoStop}%
\bibitem [{\citenamefont {Wigner}(1934)}]{Wigner34-wigner-crystal}%
  \BibitemOpen
  \bibfield  {author} {\bibinfo {author} {\bibfnamefont {E.}~\bibnamefont
  {Wigner}},\ }\bibfield  {title} {\bibinfo {title} {On the interaction of
  electrons in metals},\ }\href@noop {} {\bibfield  {journal} {\bibinfo
  {journal} {Phys. Rev.}\ }\textbf {\bibinfo {volume} {46}},\ \bibinfo {pages}
  {1002} (\bibinfo {year} {1934})}\BibitemShut {NoStop}%
\bibitem [{\citenamefont {Ung}\ \emph {et~al.}(2023)\citenamefont {Ung},
  \citenamefont {Lee},\ and\ \citenamefont
  {Reichman}}]{Ung23-wigner-crystal-moire}%
  \BibitemOpen
  \bibfield  {author} {\bibinfo {author} {\bibfnamefont {S.~F.}\ \bibnamefont
  {Ung}}, \bibinfo {author} {\bibfnamefont {J.}~\bibnamefont {Lee}},\ and\
  \bibinfo {author} {\bibfnamefont {D.~R.}\ \bibnamefont {Reichman}},\
  }\bibfield  {title} {\bibinfo {title} {Competing generalized {Wigner} crystal
  states in moir{\'e} heterostructures},\ }\href@noop {} {\bibfield  {journal}
  {\bibinfo  {journal} {{Phys. Rev. B}}\ }\textbf {\bibinfo {volume} {108}},\
  \bibinfo {pages} {245113} (\bibinfo {year} {2023})}\BibitemShut {NoStop}%
\bibitem [{\citenamefont {Tsui}\ \emph {et~al.}(2024)\citenamefont {Tsui},
  \citenamefont {He}, \citenamefont {Hu}, \citenamefont {Lake}, \citenamefont
  {Wang}, \citenamefont {Watanabe}, \citenamefont {Taniguchi}, \citenamefont
  {Zaletel},\ and\ \citenamefont {Yazdani}}]{Tsui24-wigner-crystal-expt}%
  \BibitemOpen
  \bibfield  {author} {\bibinfo {author} {\bibfnamefont {Y.-C.}\ \bibnamefont
  {Tsui}}, \bibinfo {author} {\bibfnamefont {M.}~\bibnamefont {He}}, \bibinfo
  {author} {\bibfnamefont {Y.}~\bibnamefont {Hu}}, \bibinfo {author}
  {\bibfnamefont {E.}~\bibnamefont {Lake}}, \bibinfo {author} {\bibfnamefont
  {T.}~\bibnamefont {Wang}}, \bibinfo {author} {\bibfnamefont {K.}~\bibnamefont
  {Watanabe}}, \bibinfo {author} {\bibfnamefont {T.}~\bibnamefont {Taniguchi}},
  \bibinfo {author} {\bibfnamefont {M.~P.}\ \bibnamefont {Zaletel}},\ and\
  \bibinfo {author} {\bibfnamefont {A.}~\bibnamefont {Yazdani}},\ }\bibfield
  {title} {\bibinfo {title} {Direct observation of a magnetic-field-induced
  {Wigner} crystal},\ }\href@noop {} {\bibfield  {journal} {\bibinfo  {journal}
  {Nature}\ }\textbf {\bibinfo {volume} {628}},\ \bibinfo {pages} {287}
  (\bibinfo {year} {2024})}\BibitemShut {NoStop}%
\bibitem [{\citenamefont {Gr{\"u}ner}(1988)}]{Gruner88-cdw-review}%
  \BibitemOpen
  \bibfield  {author} {\bibinfo {author} {\bibfnamefont {G.}~\bibnamefont
  {Gr{\"u}ner}},\ }\bibfield  {title} {\bibinfo {title} {The dynamics of
  charge-density waves},\ }\href@noop {} {\bibfield  {journal} {\bibinfo
  {journal} {{Rev. Mod. Phys.}}\ }\textbf {\bibinfo {volume} {60}},\ \bibinfo
  {pages} {1129} (\bibinfo {year} {1988})}\BibitemShut {NoStop}%
\bibitem [{\citenamefont {Giuliani}\ and\ \citenamefont
  {Vignale}(2008)}]{Giuliani08-quantum-liquid-book}%
  \BibitemOpen
  \bibfield  {author} {\bibinfo {author} {\bibfnamefont {G.}~\bibnamefont
  {Giuliani}}\ and\ \bibinfo {author} {\bibfnamefont {G.}~\bibnamefont
  {Vignale}},\ }\href@noop {} {\emph {\bibinfo {title} {Quantum theory of the
  electron liquid}}}\ (\bibinfo  {publisher} {Cambridge university press},\
  \bibinfo {year} {2008})\BibitemShut {NoStop}%
\bibitem [{\citenamefont {Overhauser}(1962)}]{Overhauser62-SDW}%
  \BibitemOpen
  \bibfield  {author} {\bibinfo {author} {\bibfnamefont {A.}~\bibnamefont
  {Overhauser}},\ }\bibfield  {title} {\bibinfo {title} {Spin density waves in
  an electron gas},\ }\href@noop {} {\bibfield  {journal} {\bibinfo  {journal}
  {Phys. Rev.}\ }\textbf {\bibinfo {volume} {128}},\ \bibinfo {pages} {1437}
  (\bibinfo {year} {1962})}\BibitemShut {NoStop}%
\bibitem [{\citenamefont {Rytova}(1967)}]{Rytova67-screened-coulomb}%
  \BibitemOpen
  \bibfield  {author} {\bibinfo {author} {\bibfnamefont {N.~S.}\ \bibnamefont
  {Rytova}},\ }\bibfield  {title} {\bibinfo {title} {Screened potential of a
  point charge in a thin film},\ }\href@noop {} {\bibfield  {journal} {\bibinfo
   {journal} {Mosc. Univ. Phys. Bull.}\ }\textbf {\bibinfo {volume} {3}},\
  \bibinfo {pages} {30} (\bibinfo {year} {1967})}\BibitemShut {NoStop}%
\bibitem [{\citenamefont {Keldysh}(1979)}]{Keldysh79-screened-coulomb}%
  \BibitemOpen
  \bibfield  {author} {\bibinfo {author} {\bibfnamefont {L.~V.}\ \bibnamefont
  {Keldysh}},\ }\bibfield  {title} {\bibinfo {title} {Coulomb interaction in
  thin semiconductor and semimetal films},\ }\href@noop {} {\bibfield
  {journal} {\bibinfo  {journal} {JETP Letters}\ }\textbf {\bibinfo {volume}
  {29}},\ \bibinfo {pages} {658} (\bibinfo {year} {1979})}\BibitemShut
  {NoStop}%
\bibitem [{\citenamefont {Berkelbach}\ \emph {et~al.}(2013)\citenamefont
  {Berkelbach}, \citenamefont {Hybertsen},\ and\ \citenamefont
  {Reichman}}]{Berkelbach13-dichalcogenides}%
  \BibitemOpen
  \bibfield  {author} {\bibinfo {author} {\bibfnamefont {T.~C.}\ \bibnamefont
  {Berkelbach}}, \bibinfo {author} {\bibfnamefont {M.~S.}\ \bibnamefont
  {Hybertsen}},\ and\ \bibinfo {author} {\bibfnamefont {D.~R.}\ \bibnamefont
  {Reichman}},\ }\bibfield  {title} {\bibinfo {title} {Theory of neutral and
  charged excitons in monolayer transition metal dichalcogenides},\ }\href@noop
  {} {\bibfield  {journal} {\bibinfo  {journal} {{Phys. Rev. B}}\ }\textbf
  {\bibinfo {volume} {88}},\ \bibinfo {pages} {045318} (\bibinfo {year}
  {2013})}\BibitemShut {NoStop}%
\bibitem [{\citenamefont {Monarkha}\ and\ \citenamefont
  {Syvokon}(2012)}]{Monarkha12-Wigner-crystal-review}%
  \BibitemOpen
  \bibfield  {author} {\bibinfo {author} {\bibfnamefont {Y.~P.}\ \bibnamefont
  {Monarkha}}\ and\ \bibinfo {author} {\bibfnamefont {V.}~\bibnamefont
  {Syvokon}},\ }\bibfield  {title} {\bibinfo {title} {A two-dimensional
  {Wigner} crystal},\ }\href@noop {} {\bibfield  {journal} {\bibinfo  {journal}
  {Low Temp. Phys.}\ }\textbf {\bibinfo {volume} {38}},\ \bibinfo {pages}
  {1067} (\bibinfo {year} {2012})}\BibitemShut {NoStop}%
\bibitem [{\citenamefont {Bae}\ \emph {et~al.}(2022)\citenamefont {Bae},
  \citenamefont {Wang}, \citenamefont {Scheie}, \citenamefont {Xu},
  \citenamefont {Chica}, \citenamefont {Diederich}, \citenamefont {Cenker},
  \citenamefont {Ziebel}, \citenamefont {Bai}, \citenamefont {Ren},
  \citenamefont {Dean}, \citenamefont {Delor}, \citenamefont {Xu},
  \citenamefont {Roy}, \citenamefont {Kent},\ and\ \citenamefont
  {Zhu}}]{Bae22-exciton}%
  \BibitemOpen
  \bibfield  {author} {\bibinfo {author} {\bibfnamefont {Y.~J.}\ \bibnamefont
  {Bae}}, \bibinfo {author} {\bibfnamefont {J.}~\bibnamefont {Wang}}, \bibinfo
  {author} {\bibfnamefont {A.}~\bibnamefont {Scheie}}, \bibinfo {author}
  {\bibfnamefont {J.}~\bibnamefont {Xu}}, \bibinfo {author} {\bibfnamefont
  {D.~G.}\ \bibnamefont {Chica}}, \bibinfo {author} {\bibfnamefont {G.~M.}\
  \bibnamefont {Diederich}}, \bibinfo {author} {\bibfnamefont {J.}~\bibnamefont
  {Cenker}}, \bibinfo {author} {\bibfnamefont {M.~E.}\ \bibnamefont {Ziebel}},
  \bibinfo {author} {\bibfnamefont {Y.}~\bibnamefont {Bai}}, \bibinfo {author}
  {\bibfnamefont {H.}~\bibnamefont {Ren}}, \bibinfo {author} {\bibfnamefont
  {C.~R.}\ \bibnamefont {Dean}}, \bibinfo {author} {\bibfnamefont
  {M.}~\bibnamefont {Delor}}, \bibinfo {author} {\bibfnamefont
  {X.}~\bibnamefont {Xu}}, \bibinfo {author} {\bibfnamefont {X.}~\bibnamefont
  {Roy}}, \bibinfo {author} {\bibfnamefont {A.~D.}\ \bibnamefont {Kent}},\ and\
  \bibinfo {author} {\bibfnamefont {X.}~\bibnamefont {Zhu}},\ }\bibfield
  {title} {\bibinfo {title} {Exciton-coupled coherent magnons in a {2D}
  semiconductor},\ }\href@noop {} {\bibfield  {journal} {\bibinfo  {journal}
  {Nature}\ }\textbf {\bibinfo {volume} {609}},\ \bibinfo {pages} {282}
  (\bibinfo {year} {2022})}\BibitemShut {NoStop}%
\bibitem [{\citenamefont {Wang}\ \emph {et~al.}(2023)\citenamefont {Wang},
  \citenamefont {Zhang}, \citenamefont {Yang}, \citenamefont {Lin},
  \citenamefont {Chen}, \citenamefont {Yang}, \citenamefont {Gong},
  \citenamefont {Chen}, \citenamefont {Ye},\ and\ \citenamefont
  {Liu}}]{Wang23-polariton}%
  \BibitemOpen
  \bibfield  {author} {\bibinfo {author} {\bibfnamefont {T.}~\bibnamefont
  {Wang}}, \bibinfo {author} {\bibfnamefont {D.}~\bibnamefont {Zhang}},
  \bibinfo {author} {\bibfnamefont {S.}~\bibnamefont {Yang}}, \bibinfo {author}
  {\bibfnamefont {Z.}~\bibnamefont {Lin}}, \bibinfo {author} {\bibfnamefont
  {Q.}~\bibnamefont {Chen}}, \bibinfo {author} {\bibfnamefont {J.}~\bibnamefont
  {Yang}}, \bibinfo {author} {\bibfnamefont {Q.}~\bibnamefont {Gong}}, \bibinfo
  {author} {\bibfnamefont {Z.}~\bibnamefont {Chen}}, \bibinfo {author}
  {\bibfnamefont {Y.}~\bibnamefont {Ye}},\ and\ \bibinfo {author}
  {\bibfnamefont {W.}~\bibnamefont {Liu}},\ }\bibfield  {title} {\bibinfo
  {title} {Magnetically-dressed \ce{CrSBr} exciton-polaritons in ultrastrong
  coupling regime},\ }\href@noop {} {\bibfield  {journal} {\bibinfo  {journal}
  {Nat. Commun.}\ }\textbf {\bibinfo {volume} {14}},\ \bibinfo {pages} {5966}
  (\bibinfo {year} {2023})}\BibitemShut {NoStop}%
\bibitem [{\citenamefont {Ruta}\ \emph {et~al.}(2023)\citenamefont {Ruta},
  \citenamefont {Zhang}, \citenamefont {Shao}, \citenamefont {Moore},
  \citenamefont {Acharya}, \citenamefont {Sun}, \citenamefont {Qiu},
  \citenamefont {Geurs}, \citenamefont {Kim}, \citenamefont {Fu}, \citenamefont
  {Chica}, \citenamefont {Pashov}, \citenamefont {Xu}, \citenamefont {Xiao},
  \citenamefont {Delor}, \citenamefont {Zhu}, \citenamefont {Millis},
  \citenamefont {Roy}, \citenamefont {Hone}, \citenamefont {Dean},
  \citenamefont {Katsnelson}, \citenamefont {van Schilfgaarde},\ and\
  \citenamefont {Basov}}]{Ruta23-polariton-CrSBr}%
  \BibitemOpen
  \bibfield  {author} {\bibinfo {author} {\bibfnamefont {F.~L.}\ \bibnamefont
  {Ruta}}, \bibinfo {author} {\bibfnamefont {S.}~\bibnamefont {Zhang}},
  \bibinfo {author} {\bibfnamefont {Y.}~\bibnamefont {Shao}}, \bibinfo {author}
  {\bibfnamefont {S.~L.}\ \bibnamefont {Moore}}, \bibinfo {author}
  {\bibfnamefont {S.}~\bibnamefont {Acharya}}, \bibinfo {author} {\bibfnamefont
  {Z.}~\bibnamefont {Sun}}, \bibinfo {author} {\bibfnamefont {S.}~\bibnamefont
  {Qiu}}, \bibinfo {author} {\bibfnamefont {J.}~\bibnamefont {Geurs}}, \bibinfo
  {author} {\bibfnamefont {B.~S.}\ \bibnamefont {Kim}}, \bibinfo {author}
  {\bibfnamefont {M.}~\bibnamefont {Fu}}, \bibinfo {author} {\bibfnamefont
  {D.~G.}\ \bibnamefont {Chica}}, \bibinfo {author} {\bibfnamefont
  {D.}~\bibnamefont {Pashov}}, \bibinfo {author} {\bibfnamefont
  {X.}~\bibnamefont {Xu}}, \bibinfo {author} {\bibfnamefont {D.}~\bibnamefont
  {Xiao}}, \bibinfo {author} {\bibfnamefont {M.}~\bibnamefont {Delor}},
  \bibinfo {author} {\bibfnamefont {X.-Y.}\ \bibnamefont {Zhu}}, \bibinfo
  {author} {\bibfnamefont {A.~J.}\ \bibnamefont {Millis}}, \bibinfo {author}
  {\bibfnamefont {X.}~\bibnamefont {Roy}}, \bibinfo {author} {\bibfnamefont
  {J.~C.}\ \bibnamefont {Hone}}, \bibinfo {author} {\bibfnamefont {C.~R.}\
  \bibnamefont {Dean}}, \bibinfo {author} {\bibfnamefont {M.~I.}\ \bibnamefont
  {Katsnelson}}, \bibinfo {author} {\bibfnamefont {M.}~\bibnamefont {van
  Schilfgaarde}},\ and\ \bibinfo {author} {\bibfnamefont {D.~N.}\ \bibnamefont
  {Basov}},\ }\bibfield  {title} {\bibinfo {title} {Hyperbolic exciton
  polaritons in a {van der Waals} magnet},\ }\href@noop {} {\bibfield
  {journal} {\bibinfo  {journal} {Nat. Commun.}\ }\textbf {\bibinfo {volume}
  {14}},\ \bibinfo {pages} {8261} (\bibinfo {year} {2023})}\BibitemShut
  {NoStop}%
\bibitem [{\citenamefont {Diederich}\ \emph {et~al.}(2023)\citenamefont
  {Diederich}, \citenamefont {Cenker}, \citenamefont {Ren}, \citenamefont
  {Fonseca}, \citenamefont {Chica}, \citenamefont {Bae}, \citenamefont {Zhu},
  \citenamefont {Roy}, \citenamefont {Cao}, \citenamefont {Xiao},\ and\
  \citenamefont {Xu}}]{Diederich23-exciton-magnon-CrSBr}%
  \BibitemOpen
  \bibfield  {author} {\bibinfo {author} {\bibfnamefont {G.~M.}\ \bibnamefont
  {Diederich}}, \bibinfo {author} {\bibfnamefont {J.}~\bibnamefont {Cenker}},
  \bibinfo {author} {\bibfnamefont {Y.}~\bibnamefont {Ren}}, \bibinfo {author}
  {\bibfnamefont {J.}~\bibnamefont {Fonseca}}, \bibinfo {author} {\bibfnamefont
  {D.~G.}\ \bibnamefont {Chica}}, \bibinfo {author} {\bibfnamefont {Y.~J.}\
  \bibnamefont {Bae}}, \bibinfo {author} {\bibfnamefont {X.}~\bibnamefont
  {Zhu}}, \bibinfo {author} {\bibfnamefont {X.}~\bibnamefont {Roy}}, \bibinfo
  {author} {\bibfnamefont {T.}~\bibnamefont {Cao}}, \bibinfo {author}
  {\bibfnamefont {D.}~\bibnamefont {Xiao}},\ and\ \bibinfo {author}
  {\bibfnamefont {X.}~\bibnamefont {Xu}},\ }\bibfield  {title} {\bibinfo
  {title} {Tunable interaction between excitons and hybridized magnons in a
  layered semiconductor},\ }\href@noop {} {\bibfield  {journal} {\bibinfo
  {journal} {Nat. Nanotechnol.}\ }\textbf {\bibinfo {volume} {18}},\ \bibinfo
  {pages} {23} (\bibinfo {year} {2023})}\BibitemShut {NoStop}%
\bibitem [{\citenamefont {Jiang}\ \emph {et~al.}(2024)\citenamefont {Jiang},
  \citenamefont {Li}, \citenamefont {Lyu}, \citenamefont {Xiao}, \citenamefont
  {Li}, \citenamefont {Wang}, \citenamefont {Tang}, \citenamefont {Wang},
  \citenamefont {Zhang}, \citenamefont {Liu}, \citenamefont {Yang},
  \citenamefont {Hu}, \citenamefont {Ye}, \citenamefont {Chen}, \citenamefont
  {Gao}, \citenamefont {Wu},\ and\ \citenamefont
  {Gong}}]{Jiang24-ultrafast-CrSBr}%
  \BibitemOpen
  \bibfield  {author} {\bibinfo {author} {\bibfnamefont {P.}~\bibnamefont
  {Jiang}}, \bibinfo {author} {\bibfnamefont {Y.}~\bibnamefont {Li}}, \bibinfo
  {author} {\bibfnamefont {X.}~\bibnamefont {Lyu}}, \bibinfo {author}
  {\bibfnamefont {J.}~\bibnamefont {Xiao}}, \bibinfo {author} {\bibfnamefont
  {X.}~\bibnamefont {Li}}, \bibinfo {author} {\bibfnamefont {T.}~\bibnamefont
  {Wang}}, \bibinfo {author} {\bibfnamefont {J.}~\bibnamefont {Tang}}, \bibinfo
  {author} {\bibfnamefont {Y.}~\bibnamefont {Wang}}, \bibinfo {author}
  {\bibfnamefont {L.}~\bibnamefont {Zhang}}, \bibinfo {author} {\bibfnamefont
  {Y.}~\bibnamefont {Liu}}, \bibinfo {author} {\bibfnamefont {H.}~\bibnamefont
  {Yang}}, \bibinfo {author} {\bibfnamefont {X.}~\bibnamefont {Hu}}, \bibinfo
  {author} {\bibfnamefont {Y.}~\bibnamefont {Ye}}, \bibinfo {author}
  {\bibfnamefont {Z.}~\bibnamefont {Chen}}, \bibinfo {author} {\bibfnamefont
  {Y.}~\bibnamefont {Gao}}, \bibinfo {author} {\bibfnamefont {C.}~\bibnamefont
  {Wu}},\ and\ \bibinfo {author} {\bibfnamefont {Q.}~\bibnamefont {Gong}},\
  }\bibfield  {title} {\bibinfo {title} {Ultrafast electron dynamics dominated
  by electron–phonon coupling in {CrSBr} revealed by photoemission electron
  microscopy},\ }\href {https://doi.org/10.1021/acs.jpcc.4c07231} {\bibfield
  {journal} {\bibinfo  {journal} {J. Phys. Chem. C}\ }\textbf {\bibinfo
  {volume} {128}},\ \bibinfo {pages} {21855} (\bibinfo {year}
  {2024})}\BibitemShut {NoStop}%
\bibitem [{\citenamefont {Datta}\ \emph {et~al.}(2024)\citenamefont {Datta},
  \citenamefont {Adak}, \citenamefont {Yu}, \citenamefont {Dharmapalan},
  \citenamefont {Hall}, \citenamefont {Vakulenko}, \citenamefont
  {Komissarenko}, \citenamefont {Kurganov}, \citenamefont {Quan}, \citenamefont
  {Wang}, \citenamefont {Mosina}, \citenamefont {Sofer}, \citenamefont
  {Pashov}, \citenamefont {van Schilfgaarde}, \citenamefont {Acharya},
  \citenamefont {Kamra}, \citenamefont {Sfeir}, \citenamefont {Al\`u},
  \citenamefont {Khanikaev},\ and\ \citenamefont
  {Menon}}]{Datta24-CrSBr-magnon}%
  \BibitemOpen
  \bibfield  {author} {\bibinfo {author} {\bibfnamefont {B.}~\bibnamefont
  {Datta}}, \bibinfo {author} {\bibfnamefont {P.~C.}\ \bibnamefont {Adak}},
  \bibinfo {author} {\bibfnamefont {S.}~\bibnamefont {Yu}}, \bibinfo {author}
  {\bibfnamefont {A.~V.}\ \bibnamefont {Dharmapalan}}, \bibinfo {author}
  {\bibfnamefont {S.~J.}\ \bibnamefont {Hall}}, \bibinfo {author}
  {\bibfnamefont {A.}~\bibnamefont {Vakulenko}}, \bibinfo {author}
  {\bibfnamefont {F.}~\bibnamefont {Komissarenko}}, \bibinfo {author}
  {\bibfnamefont {E.}~\bibnamefont {Kurganov}}, \bibinfo {author}
  {\bibfnamefont {J.}~\bibnamefont {Quan}}, \bibinfo {author} {\bibfnamefont
  {W.}~\bibnamefont {Wang}}, \bibinfo {author} {\bibfnamefont {K.}~\bibnamefont
  {Mosina}}, \bibinfo {author} {\bibfnamefont {Z.}~\bibnamefont {Sofer}},
  \bibinfo {author} {\bibfnamefont {D.}~\bibnamefont {Pashov}}, \bibinfo
  {author} {\bibfnamefont {M.}~\bibnamefont {van Schilfgaarde}}, \bibinfo
  {author} {\bibfnamefont {S.}~\bibnamefont {Acharya}}, \bibinfo {author}
  {\bibfnamefont {A.}~\bibnamefont {Kamra}}, \bibinfo {author} {\bibfnamefont
  {M.~Y.}\ \bibnamefont {Sfeir}}, \bibinfo {author} {\bibfnamefont
  {A.}~\bibnamefont {Al\`u}}, \bibinfo {author} {\bibfnamefont {A.~B.}\
  \bibnamefont {Khanikaev}},\ and\ \bibinfo {author} {\bibfnamefont {V.~M.}\
  \bibnamefont {Menon}},\ }\bibfield  {title} {\bibinfo {title}
  {Magnon-mediated exciton-exciton interaction in a {van der Waals}
  antiferromagnet},\ }\href@noop {} {\bibfield  {journal} {\bibinfo  {journal}
  {arXiv preprint arXiv:2409.18501}\ } (\bibinfo {year} {2024})}\BibitemShut
  {NoStop}%
\end{thebibliography}%

\end{document}